\documentclass{aa}
\usepackage{graphicx}
\usepackage{txfonts}
\begin{document}

\def\simlt{\mathrel{\rlap{\lower 3pt\hbox{$\sim$}}\raise 2.0pt\hbox{$<$}}}
\def\simgt{\mathrel{\rlap{\lower 3pt\hbox{$\sim$}}\raise 2.0pt\hbox{$>$}}}

   \title{The LOFAR Two-metre Sky Survey Deep fields: \\
   A new analysis of low-frequency radio luminosity as a star-formation tracer in the Lockman Hole region}

   \titlerunning{A new analysis of low-frequency radio luminosity as a star-formation tracer in the Lockman Hole region}

   \author{M. Bonato
          \inst{1,2,3}
          \and
          I. Prandoni\inst{1}
          \and
          G. De Zotti\inst{3}
          \and
           P. N. Best\inst{4}
          \and
          M. Bondi\inst{1}
          \and
          G. Calistro Rivera\inst{5}
          \and
          R. K. Cochrane\inst{6}
          \and
          G. G{\"u}rkan\inst{7}
          \and
          P. Haskell\inst{8}
          \and
          R. Kondapally\inst{4}
          \and
          M. Magliocchetti\inst{9}
          \and
          S. K. Leslie\inst{10}
          \and
          K. Malek\inst{11,12}
          \and
          H. J. A. R{\"o}ttgering\inst{10}
          \and
          D. J. B. Smith\inst{8}
          \and
          C. Tasse\inst{13,14}
          \and
          L. Wang\inst{15,16}}

   \institute{INAF$-$Istituto di Radioastronomia, Via Gobetti 101, I-40129, Bologna, Italy
         \email{matteo.bonato@inaf.it}
         \and
         Italian ALMA Regional Centre, Via Gobetti 101, I-40129, Bologna, Italy
         \and
         INAF$-$Osservatorio Astronomico di Padova, Vicolo dell'Osservatorio 5, I-35122, Padova, Italy
         \and
         SUPA, Institute for Astronomy, Royal Observatory, Blackford Hill, Edinburgh, EH9 3HJ, UK
         \and
         European Southern Observatory, Karl$-$Schwarzchild$-$Strasse 2, 85748, Garching bei München, Germany
         \and
         Harvard-Smithsonian Center for Astrophysics, 60 Garden St, Cambridge, MA 02138, USA
         \and
         Th\"uringer Landessternwarte, Sternwarte 5, D-07778 Tautenburg, Germany
         \and
        Centre for Astrophysics Research, University of Hertfordshire, Hatfield, AL10 9AB, UK
         \and
         INAF$-$Istituto di Astrofisica e Planetologia Spaziali, Via del Fosso del Cavaliere 100, 00133 Roma, Italy
         \and
         Leiden Observatory, Leiden University, PO Box 9513, NL-2300 RA Leiden, The Netherlands
         \and
         National Centre for Nuclear Research, ul. Pasteura 7, 02-093 Warszawa, Poland
         \and
         Aix Marseille Univ. CNRS, CNES, LAM, Marseille, France
         \and
         GEPI \& USN, Observatoire de Paris, CNRS, Universit\'e Paris Diderot, 5 place Jules Janssen, 92190 Meudon, France
         \and
         Centre for Radio Astronomy Techniques and Technologies, Department of Physics and Electronics, Rhodes University, Grahamstown 6140, South Africa
         \and
         SRON Netherlands Institute for Space Research, Landleven 12, 9747 AD, Groningen, The Netherlands
         \and
         Kapteyn Astronomical Institute, University of Groningen, Postbus 800, 9700 AV Groningen, the Netherlands
             }

   \date{Received XXXXXXXX; accepted XXXXXXXX}


  \abstract
{We have exploited LOFAR deep observations of the Lockman Hole field at 150\,MHz to investigate the relation between the radio luminosity of star-forming galaxies (SFGs) and their star formation rates (SFRs), as well as its dependence on stellar mass and redshift. The adopted source classification, SFRs and stellar masses are consensus estimates based on a combination of four different SED fitting methods. We note a flattening of radio spectra of a substantial minority of sources below $\sim 1.4\,$GHz. Such sources have thus a lower ``radio-loudness'' level at 150\,MHz than expected from extrapolations from $1.4\,$GHz using the average spectral index.
We found a weak trend towards a lower $\hbox{SFR}/L_{150\,\rm MHz}$ ratio for higher stellar mass,
$M_\star$. We argue that such a trend may account for most of the apparent redshift evolution of the $L_{150\,\rm MHz}/\hbox{SFR}$ ratio, in line with previous work. Our data indicate a weaker evolution than found by some previous analyses. We also find a weaker evolution with redshift of the specific star formation rate than found by several (but not all) previous studies. Our radio selection provides a view of the distribution of galaxies in the $\hbox{SFR}$--$M_\star$ plane complementary to that of optical/near-IR selection. It suggests a higher uniformity of the star formation history of galaxies than implied by some analyses of optical/near-IR data. We have derived luminosity functions at 150\,MHz of both SFGs and radio-quiet (RQ) AGN at
various redshifts. Our results are in very good agreement with the T-RECS simulations and with literature estimates.
We also present explicit estimates of SFR
functions of SFGs and RQ AGN at several redshifts derived from our radio survey data. }


   \maketitle
%

\section{Introduction}\label{sec:introduction}
The fact that radio-source counts are increasingly dominated by
star-forming galaxies (SFGs) at 1.4\,GHz flux densities fainter than a few hundred $\mu$Jy
\citep{Padovani2015, Smolcic2017b, Prandoni2018, RetanaMontenegro2018, Algera2020, vanderVlugt2021} has opened a new era in radio astronomy with a growing role of deep radio surveys in the study of the star-formation history of
galaxies (see \citealt{DeZotti2010} and \citealt{Padovani2016} for reviews).

The huge increase in sensitivity of the upgraded radio interferometers such as the Karl G. Jansky Very Large
Array (JVLA) and of the new generation of radio telescopes (e.g. the South African MeerKAT, the Australian Square Kilometre Array Pathfinder (ASKAP) and the Low-Frequency Array (LOFAR)) has
offered the possibility of exploiting radio surveys as a probe of galaxy
evolution up to high redshifts. While rest-frame ultraviolet (UV) surveys have allowed the
investigation of the galaxy luminosity functions up to $z\sim 10$
\citep{Oesch2018, Bouwens2019, Bowler2020}, they miss a lot of dust-obscured
star formation and may be contaminated by emission from active galactic nuclei (AGN). On the other hand, far-infrared (FIR)/sub-millimeter (sub-mm) surveys measure only starlight
reprocessed by dust which may include the contribution of evolved stars.  Also large-area FIR/sub-mm surveys currently suffer from resolution limitations implying severe confusion limits (cf., e.g., \textit{Herschel}/SPIRE surveys).  Radio emission has the advantages of being dust--independent and powered by recent star
formation, although it may also be contaminated by radio AGN. The new radio surveys have the additional assets of large increases in sensitivity, resolution and survey speed.

The use of radio luminosity as a star-formation rate (SFR) indicator hinges
upon its tight correlation with the FIR luminosity, which is a well 
established measure of dust-enshrouded star formation \citep[e.g.,][]{Helou1985,
Condon1992}. The continuum emission of star-forming galaxies at low radio frequencies ($\lesssim$1\,GHz) is dominated by
synchrotron radiation from relativistic electrons mostly accelerated by
supernova remnants, produced by short-lived massive stars \citep{Condon1992, Murphy2011}. There is therefore a
clear link between radio emission and SFR. However the quantitative relation
between these two quantities depends on poorly understood physical processes and on several parameters (magnetic field intensity, energy density of the radiation field, cooling and confinement of relativistic electrons, among others) so
that it cannot be derived from first principles.

Empirical determinations of the radio luminosity--SFR calibration are exposed to several difficulties, as detailed below.  One such issue is selection effects: the radio selection favours higher $L_{\rm radio}/\hbox{SFR}$ ratios while the opposite is true for the FIR/sub-mm selection which favours high SFRs. The bias associated with optical/near-IR selection is less straightforward; however such selection under-represents dust-enshrouded, high-SFR galaxies. Another issue is the possibility that other parameters such as the stellar mass or the redshift have a role. Moreover, different approaches and algorithms have been used to derive the SFRs and the quality of the used data is uneven. A minor, but still significant effect is due to errors in extrapolations of relations derived at different radio frequencies. It is therefore not surprising that varying results are obtained.

Several studies reported a constant radio luminosity to SFR ratio over several orders of
magnitude in luminosity \citep[e.g.,][]{Yun2001, Moric2010, Murphy2011, Murphy2012,
CalistroRivera2017, Delhaize2017, Solarz2019, Wang2019}. These studies
estimated the SFR either via spectral energy distribution (SED) fitting or from the IR
luminosity. However, for the same initial mass function (IMF) there are substantial differences among the reported calibrations of SFRs. For a \citet{Chabrier2003} IMF the values of $\log(L_{1.4}/\hbox{W}\,\hbox{Hz}^{-1})-\log(\hbox{SFR}/M_\odot\,\hbox{yr}^{-1})$
at $z=0$ range from $20.96\pm 0.03$ \citep{Delhaize2017} to $21.39\pm 0.04$ \citep{CalistroRivera2017}, corresponding to a factor of $\simeq 2.7$ difference in the local $L_{1.4}$/SFR ratio.

Both theoretical \citep{ChiWolfendale1990, Lacki2010} and
empirical \citep{Bell2003, Massardi2010, Mancuso2015b, Bonato2017} arguments
point to lower $L_{1.4}$/SFR ratios for dwarf galaxies from which relativistic electrons can escape before losing most of their energy via synchrotron emission. Observational evidence
of super-linear radio/IR or radio/SFR relations,
$L_{\rm radio}\propto L_{\rm IR}^\delta$ or $L_{\rm radio}\propto
\hbox{SFR}^\delta$ with $\delta > 1$, has been reported by several studies
\citep{Klein1984, DevereuxEales1989, PriceDuric1992, Hodge2008, Basu2015,
Davies2017, Brown2017, Wang2019, Molnar2021}.

On the other hand, \citet{Gurkan2018} found evidence for excess radio emission
from SFGs with $\hbox{SFR}\lesssim 1\,M_\odot\,\hbox{yr}^{-1}$. These authors
also found that the $L_{150\,\rm MHz}$/SFR ratio increases with stellar mass. Compelling evidence of larger $L_{150\,\rm MHz}$/SFR ratios for more massive galaxies has been reported by \citet{Smith2020} using the most sensitive 150\,MHz data in existence. A weaker but highly significant mass dependence of the $L_{1.4}$/SFR ratio has been
reported also by \citet{Delvecchio2020}.

Another debated issue is the redshift dependence of the $L_{\rm radio}$--SFR
relation. A decrease of the synchrotron luminosity at fixed SFR at high $z$ was
predicted \citep{Murphy2009, LackiThompson2010, Schober2016} as a consequence of the higher
inverse Compton losses of relativistic electrons off the cosmic microwave
background whose energy density rapidly increases with $z$. Contrary to these
predictions, a slight but significant \textit{increase} of the $L_{\rm
radio}$/SFR ratio with $z$ was reported by several studies \citep{Ivison2010b,
Casey2012, Magnelli2015, CalistroRivera2017, Delhaize2017, Delvecchio2020}. The
redshift dependence is controversial, however. Other works did not find
significant evidence for the evolution of the $L_{\rm radio}/$SFR ratio
\citep{Ibar2008, Garn2009, Mao2011, Bourne2011, Smith2014, Duncan2020, Smith2020}.
\citet{Sargent2010a, Sargent2010b} pointed out selection biases that may yield
apparent evolution \citep[see also][]{Smith2020}. 

\citet{Molnar2018} observed a significant difference between the redshift evolution of the $L_{\rm
radio}$/SFR ratio between spheroidal and disc-dominated SFGs, with the
latter showing very little variations. They argued that the observed evolution
of spheroidal SFGs might be ascribed to some residual AGN activity. \citet{Smith2020} pointed out that the dependence of radio emission on stellar mass may account for the redshift evolution of the $L_{1.4}$/SFR ratio found by \citet{CalistroRivera2017} and \citet{Delhaize2017}; this is because the higher-$z$ galaxies in the samples tend to be more massive due to selection biases. \citet{Molnar2021} also argued that the apparent redshift evolution reported at GHz frequencies can be due
to a selection effect, i.e. to a redshift-dependent sampling of different parts of a non-linear FIR/SFR relation.

The multifaceted uncertainties mentioned above may undermine the use of radio
continuum luminosity as a SFR tracer unless the details of the $L_{\rm
radio}$--SFR relation are settled.

This paper aims at providing a contribution in this direction, by exploiting deep observations with the Low Frequency Array (LOFAR). LOFAR is carrying out a sensitive 120$-$168\,MHz survey of the entire northern sky, the LOFAR Two$-$metre Sky Survey (LoTSS). \citet{Shimwell2017, Shimwell2019} presented the first data release (LoTSS$-$DR1), which covers  424\,deg${^2}$ with a median sensitivity S$_{144\,\rm MHz}\sim$71$\mu$Jy/beam. LoTSS also includes  deeper observations of well$-$known extra$-$galactic regions where an exceptional wealth of multi-band datasets are available. The goal is to eventually reach rms sensitivities of $\sim$10\,$\mu$Jy/beam in at least some of these regions (\citealt{Rottgering2011}). The first data release of the LoTSS Deep Fields includes the following
regions: the Lockman Hole (LH),  the Bo\"{o}tes and  the European Large$-$Area ISO Survey-North 1 (ELAIS-N1) fields. For each field, the released LOFAR data cover an area of $\sim$25\,deg${^2}$  (corresponding to one LOFAR pointing), with median sensitivities ranging from $\sim 20$ to $\sim 40\,\mu$Jy/beam in the central $\sim 10$\,deg${^2}$ (\citealt{Tasse2020}; \citealt{Sabater2020}, papers I and II of this series). Deep and wide optical and IR data are available over several square-degree areas in each of LoTSS Deep Fields. Over these common sub-regions, an extensive process of multi-wavelength cross-matching and source characterisation was carried out. This yielded cleaner and more reliable radio source catalogues. The images were pixel-matched in each waveband and aperture-matched forced photometry, from ultraviolet to infrared wavelengths, was extracted. The process is extensively described by \citet{Kondapally2020}, paper III of this series. This allowed us to derive high-quality photometric redshifts for around 5 million objects across the three fields (see \citealt{Duncan2020}, paper IV of this series, for more details). Finally SED fitting techniques were exploited to provide accurate luminosities, SFRs and stellar masses of the radio-detected galaxies \citep[][paper V]{Best2020}. The radio and value--added  catalogues, are available on the LOFAR Surveys Data Release site web-page: \url{http://www.lofar-surveys/releases.html}.

A detailed analysis of the SFR--$L_{150\,\rm MHz}$ relation based on LoTSS Deep Fields data was presented by \citet{Smith2020}. That study was based on an IRAC-selected sample in the ELAIS-N1 field. In this paper we  instead focus our analysis on LOFAR-detected sources in the  LH \citep{Lockman1986} and look at somewhat different topics with respect to \citet{Smith2020}, including luminosity functions at 150\,MHz and SFR functions. Choosing this field also allows us to take advantage of the deep survey at 1.4\,GHz carried out by \citet{Prandoni2018} with the Westerbork Synthesis Radio Telescope (WSRT) to investigate the effect of different selection frequencies on the source classification.  The WSRT data were analysed by \citet{Bonato2020}.


The layout of the paper is the following. Section~\ref{sec:data_class} presents a
short description of the sample and of the source classification; details are
given in other papers of this series. The basic properties of the source
populations are described in Sect.~\ref{sec:results}. In particular, we compare the source classification based on LOFAR data with the independent classification by \citet{Bonato2020} based on 1.4\,GHz WSRT data, discuss the relationship between the radio luminosity of SFGs and their SFR as well as between their SFR and their stellar mass, and present new estimates of radio luminosity functions and of SFR functions at several redshifts.  The main conclusions are
summarized in Sect.~\ref{sec:conclusions}.

Throughout this paper we use, for the radio flux density, the convention $S_\nu\propto \nu^\alpha$ and adopt
a flat $\Lambda \rm CDM$ cosmology with $\Omega_{\rm m} = 0.31$,
$\Omega_{\Lambda} = 0.69$ and $h=H_0/100\, \rm km\,s^{-1}\,Mpc^{-1} = 0.677$
\citep{Planck2020parameters}.


\begin{figure*}
\begin{center}
\includegraphics[width=0.49\textwidth]{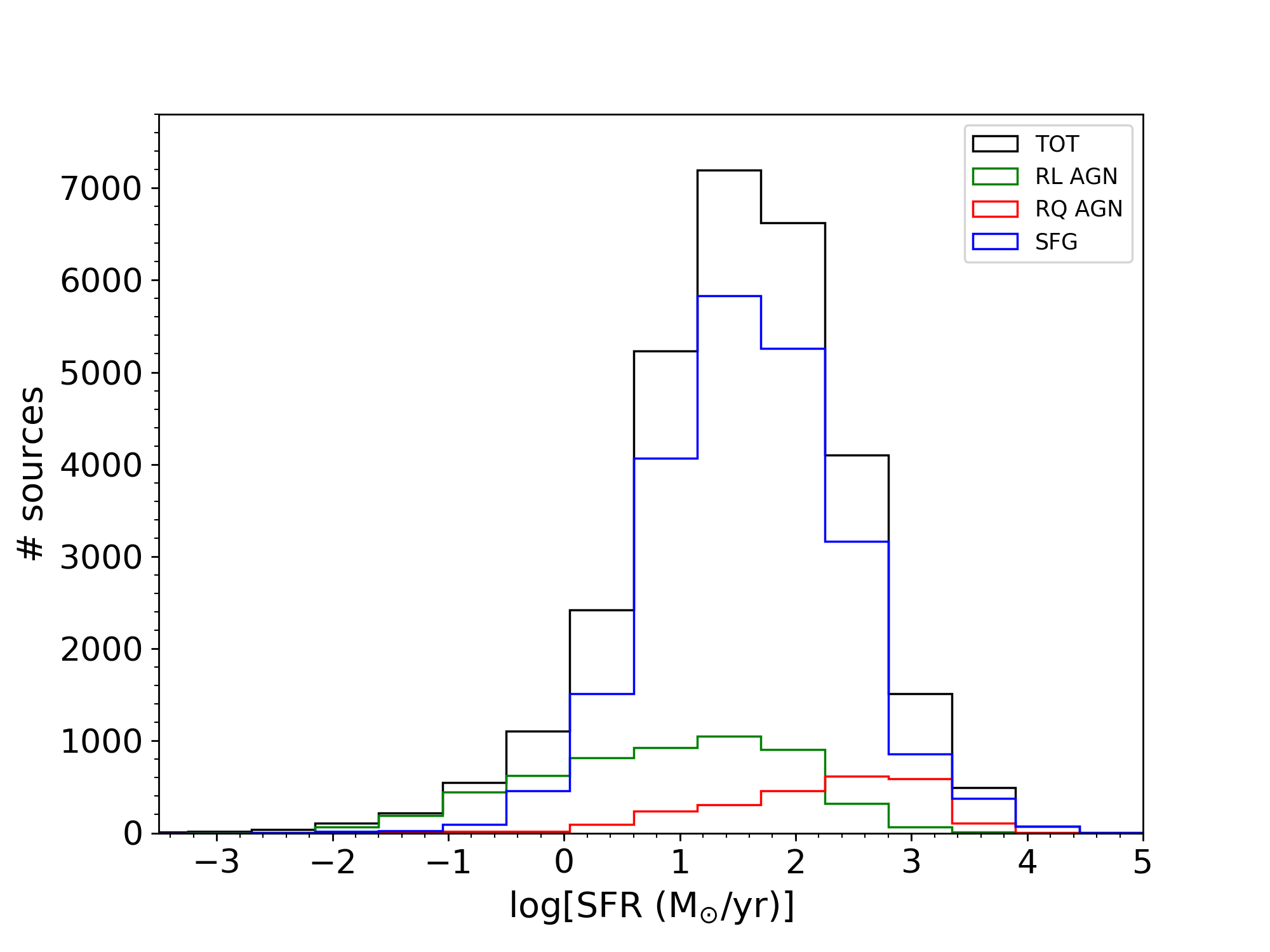}
\includegraphics[width=0.49\textwidth]{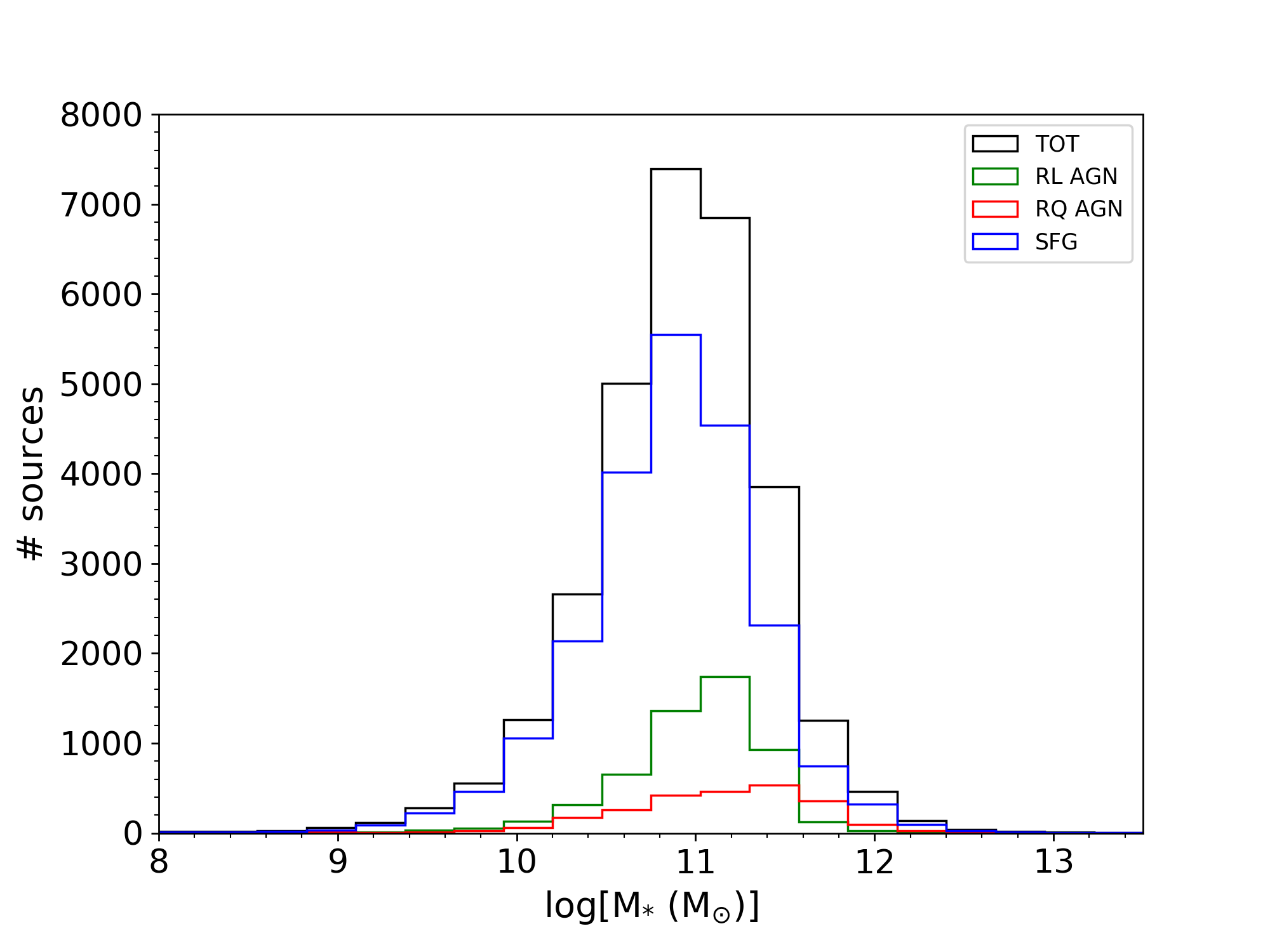}
\caption{Distributions of SFRs (left) and of stellar masses (right) of source populations
 in our sample, subdivided into SFGs, RL and RQ AGN. The distribution of SFRs of RL AGN extends to very low values,
 consistent with the fact that their radio emission is unrelated to the SFR; especially in the nearby universe
 they are generally associated to passive ellipticals. Both RQ and RL AGN preferentially reside in massive galaxies,
 with median stellar masses significantly higher than the mean stellar mass of SFGs. See text for more detailed comments.  }
 \label{fig:SFR_distributions}
  \end{center}
\end{figure*}

\begin{figure}
\begin{center}
\includegraphics[width=0.49\textwidth]{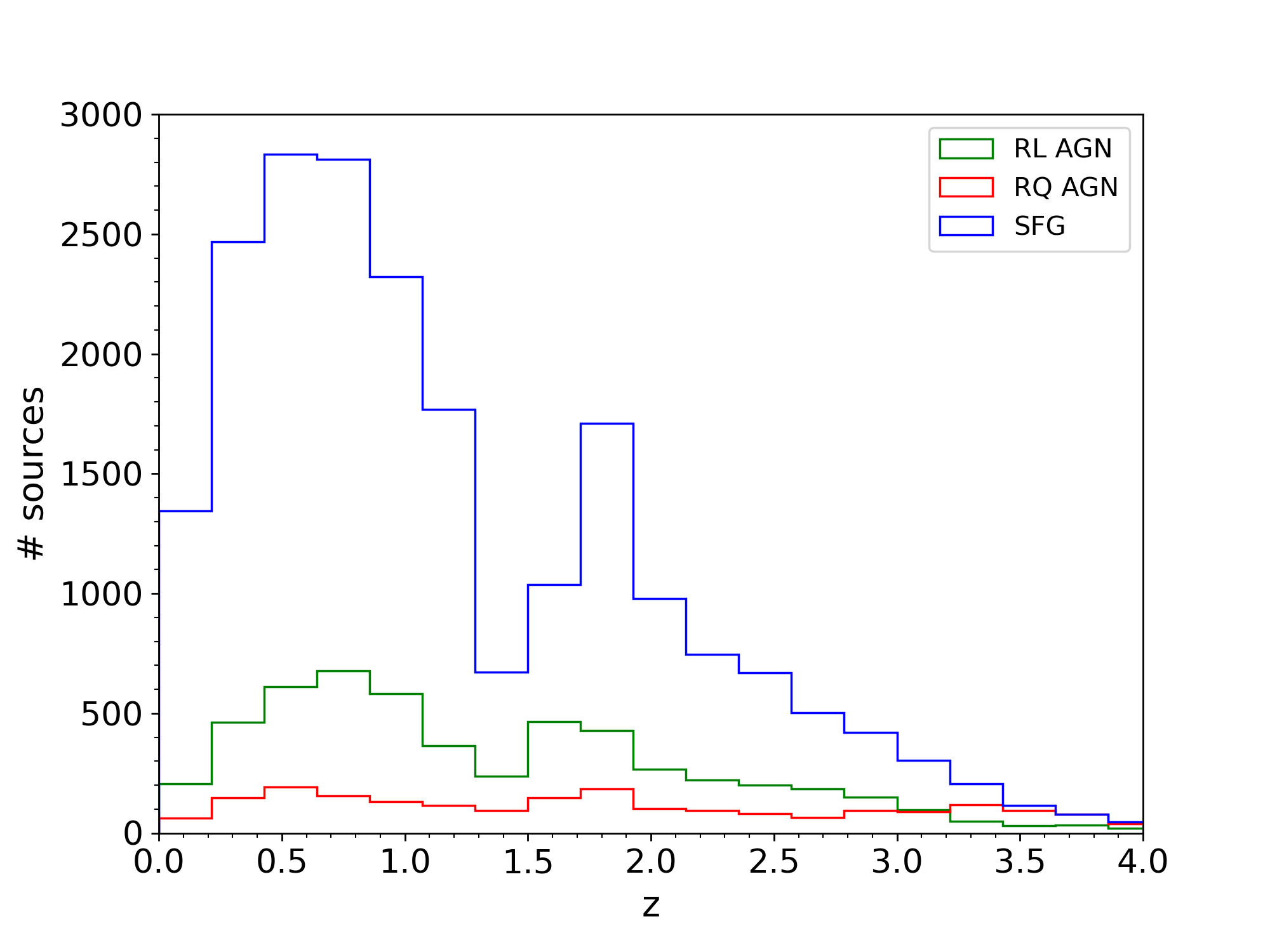}
\caption{Redshift distributions of sources in our sample, subdivided into SFGs, RL and RQ AGN. Most redshifts, especially at $z> 1$, are photometric. Their quality deteriorates for $z> 1.5$ and the distributions at $z> 4$ are unreliable. We have therefore cut them off at $z=4$. See text for a comment on the origin of the dip at $z\simeq 1.3$.  }
 \label{fig:z_distributions}
  \end{center}
\end{figure}

\section{Data and classification}\label{sec:data_class}

\subsection{Data}\label{sect:data}

The first release of the LoTSS Deep Fields data included LOFAR observations of the LH totalling $\sim$112\,h. The data were calibrated and imaged taking into account direction$-$dependent effects as described in paper I. The frequency band used to produce the LH radio image is $\sim$120\,MHz to $\sim$168\,MHz (corresponding to a central frequency of $\sim$144\,MHz). The median rms sensitivity over the fully imaged field of view (25$\,\hbox{deg}^2$) is 40\,$\mu$Jy/beam \citep{Mandal2020} at a resolution of $6''$ (full-width at half-maximum of the restoring beam). Over this area, a catalogue of 50,112 radio sources was extracted using the Python Blob Detector and Source Finder (PyBDSF; \citealt{Mohan2015}) down to a peak flux density detection threshold of 5$\sigma$, where $\sigma$ is the local rms noise (see paper I for more details on the source extraction).

Deep multi-frequency data are available for part of the field. Actually, the LH is the LoTSS field with the largest area of multi-wavelength coverage, as shown by the footprint in Fig.\,1 of \citet{Kondapally2020}. The optical data come from the \textit{Spitzer} Adaptation of the Red-sequence Cluster Survey \citep[SpARCS;][]{Wilson2009, Muzzin2009} and from the Red
Cluster Sequence Lensing Survey \citep[RCSLenS;][]{Hildebrandt2016}. The SpARCS data consist of images in the $u$, $g$, $r$ and $z$ filters, covering $13.32\,\hbox{deg}^2$ of the field. The RCSLenS observations cover $16.63\,\hbox{deg}^2$ in the $g$, $r$, $i$ and $z$ filters, but there are gaps between pointings.

The near-- and far--UV (NUV and FUV) data come from the Release 6 and 7 of the Deep Imaging Survey (DIS) with the Galaxy Evolution Explorer (GALEX) space telescope \citep{Martin2005}. The near-IR data ($J$ and $K$ bands) are taken from the UK Infrared Deep Sky Survey (UKIDSS) Deep Extragalactic Survey (DXS) DR10 \citep{Lawrence2007} which covers about $8.16\,\hbox{deg}^2$ of the field. Around $11\,\hbox{deg}^2$ of the LH field were covered by the \textit{Spitzer} Wide-area Infra-Red Extragalactic (SWIRE) survey \citep{Lonsdale2003} in all 4 InfraRed Array Camera (IRAC) channels (at 3.6, 4.5, 5.8 and $8.0\,\mu$m). A smaller area ($\simeq5.6\,\hbox{deg}^2$) was also covered by the \textit{Spitzer} Extragalactic Representative
Volume Survey \citep[SERVS;][]{Mauduit2012}, about one magnitude deeper than the SWIRE survey but in only two channels (3.6 and $4.5\,\mu$m; it was carried out during the \textit{Spitzer} warm mission).

The LH is also part of the \textit{Herschel} Multi-tiered Extragalactic Survey \citep[HerMES;][]{Oliver2012} and was observed with the Multiband Imaging Photometer for \textit{Spitzer} \citep[MIPS;][]{Rieke2004}. The HerMES data include photometry with the Spectral and Photometric Imaging Receiver \citep[SPIRE;][]{Griffin2010} at 250, 350 and $500\,\mu$m, and with the Photodetector Array Camera and Spectrometer \citep[PACS;][]{Poglitsch2010} at 100 and $160\,\mu$m. As for MIPS, only $24\,\mu$m data were used. The $70\,\mu$m data from both MIPS or PACS were ignored due to their poorer sensitivity.

It is clear from the above summary that the radio data cover a larger area than the ancillary multi-wavelength data. The cross-matching was performed in the central region of the radio LH field, covered by SpARCS $r$-band and SWIRE surveys. It is indicated by the shaded light blue region in the footprint in the central panel of Fig.\,1 of \citet{Kondapally2020} where  the coverage of the main surveys used is shown. This region has an area of $\approx 10.28\,\hbox{deg}^2$ after masking regions around bright stars. The median rms sensitivity over this area is $\sim$29\,$\mu$Jy/beam (\citealt{Mandal2020}). In total, data for 16 bands were collected \citep[cf. Table~B.3 of][]{Duncan2020}, although only $\simeq 41\%$ of our sources were imaged in all bands.

The optical identification of radio sources was made using a combination of statistical cross-matching and visual analysis. The colour-based adaptation of the Likelihood Ratio (LR) method developed by \citet{Nisbet2018} was used for automated, statistical cross identifications  for sources where the radio position provided an accurate measurement of the expected position of the optical/IR host galaxy \citep[see also][]{Williams2019}. A decision tree was constructed to identify which sources met this condition. Whenever the decision tree indicated that the LR method was not suitable, the identification of counterparts was made by visual classification. The method, described in detail by \citet{Kondapally2020}, allowed us to associate multiple components of extended radio sources and to deblend sources nearby in projection.

The final LH cross-matched catalogue contains 31,162 sources. A few numbers to characterize the sample: only 522 sources ($\approx 1.7\%$) have multiple non-associated radio components; 30,402 ($\approx 97.6\%$) are optically identified
and 30,207 ($\approx 96.9\%$) have either spectroscopic (1,466) or high-quality
photometric redshifts (28,741) based on at least five optical bands. Paper III argued that most of the unidentified radio sources are likely
obscured AGN at $z >3$.

\citet{Mandal2020} present a detailed investigation of  the completeness of the catalogue taking into account the resolution bias (i.e. the fact that extended sources are more easily missed in catalogues at low SNRs than point sources), as well as of other biases (like e.g. the Eddington bias; \citealt{Eddington1913}). They provide flux density-dependent correction factors by which each source should be weighted to properly account for incompleteness effects. Such weights include the effect of varying sensitivity across the field, i.e. the fact that the effective area of the catalogue is a function of the source flux density. \citet{Mandal2020} also evaluate the systematic uncertainties in these corrections for $\ge 5\,\sigma$ detections with a median rms noise $\sigma = 31\,\mu\rm Jy\,\hbox{beam}^{-1}$. We will use the results of this investigation for our analysis.

\subsection{Classification}\label{sect:classification}

\citet[][paper V]{Best2020} describe the method used to classify the optically identified radio sources.
Four different SED fitting methods, optimized for different kinds of sources, were used for source classification and to
estimate the SFR and the stellar mass, $M_\star$, of the host galaxies: MAGPHYS
\citep{daCunha2008}, BAGPIPES \citep{Carnall2018}, CIGALE \citep{Boquien2019}
and AGNfitter \citep{CalistroRivera2016}. CIGALE was run with AGN models by
both \citet{Fritz2006} and \citet{Stalevski2012,Stalevski2016}. We refer to paper V for full details.

As input to the SED fitting, imaging in all 16\,bands (cf. sub-sect.~\ref{sect:data}) was available for about 41\% of our sources; in at least 14 bands for about 65\% of sources (the missing bands were mostly those at 5.8 and $8.0\,\mu$m); in at least 12 bands for about 90\% of sources; in at least 10 bands for about 94\% of sources; 96.9\% of sources have imaging in at least 5 (optical/near-IR) bands. Even non-detections were used in the SED fitting.


\citet{Best2020} used the outputs of the four different SED fitting algorithms to classify the radio galaxies. Objects for which the SED fits showed evidence of nuclear activity were classified as AGN: these exhibited either significant AGN fractions determined by the CIGALE and/or AGNfitter SED fits, and/or the CIGALE/AGNfitter routines were able to provide much superior fits to the SED than the MAGPHYS and BAGPIPES algorithms that did not include an AGN component.  A small number of objects classified as AGN based on their X-ray emission or existing optical spectroscopy were added to this sample.

\citet{Best2020}  showed that, for the SFGs, MAGPHYS and BAGPIPES yielded better fits and more reliable estimates of SFRs and stellar masses.  The consensus values for SFRs and stellar masses, $M_\star$, are the logarithmic average of the results from these fits unless either is a bad fit; in this case the other method was adopted. If neither method provided an acceptable fit, SFR and $M_\star$ could not be derived. This happened for just under 5\% of sources (of which 3\% were the cases where no photo-$z$ was available and so no fitting was possible, and 2\% were cases where the fitting was unreliable). Secure AGN SEDs are much better fitted by CIGALE or AGNfitter.
For SFRs and $M_\star$ the average of the two CIGALE fits was taken. Again, if any of the fits were bad, they were not used.

Using a determined  consensus SFR, objects were identified as radio-excess sources if they showed more than 0.7 dex excess of radio emission over the peak of the population at that SFR. If AGN also show a radio excess they were classified as high excitation radio galaxies (HERG), otherwise as radio-quiet (RQ) AGN. Sources with no detectable AGN signatures in the SED, but showing excess radio emission were classified as low-excitation radio galaxies (LERG)\footnote{Spectrophotometry of radio AGN has highlighted a fundamental dichotomy leading to the definition of two populations: high-- and low- excitation radio galaxies \citep[HERG and LERG, respectively;][]{Laing1994}. The two populations show differences in accretion, evolution and host galaxy properties \citep{BestHeckman2012}. HERG have higher accretion rates, in units of the Eddington rate, than LERG, consistent with the dichotomy being due to a switch from a radiatively efficient to a radiatively inefficient accretion mode. Also HERG show a strong cosmological evolution of the luminosity function while LERG evolve slowly if at all. Moreover, HERG are typically associated to lower black hole masses and are hosted by galaxies with lower stellar masses and younger stellar populations.}. The other sources were classified as SFGs.  Sources that showed clear evidence for very extended radio emission (i.e. jets) were also classified as radio-excess based on their radio properties. We refer to paper V for a thorough description of the process. In the following, LERG and HERG are considered together as a single class of radio-loud (RL) AGN.


Stellar mass  and SFR determinations are available for
31,012 and 29,828 sources, respectively. The distributions of SFRs and of
$M_\star$ for each population, as well as the total are shown in
Fig.\,\ref{fig:SFR_distributions}. We have checked that the numbers of sources in all stellar mass and
redshift bins are fully consistent with observational estimates of redshift-dependent galaxy stellar mass functions derived for LoTSS deep fields and for other fields in the literature \citep[see Fig.~12 of][]{Duncan2020}.

RL AGN have the broadest distribution of
SFRs, with a substantial tail extending to very low values, as expected since, especially at low $z$, they are frequently associated to evolved early-type galaxies; their median is $\log({\rm
SFR}/M_\odot\,\hbox{yr}^{-1})=0.92$ with standard deviation $\sigma_{\log({\rm
SFR})}=1.46$. On the contrary, the SFR distribution of RQ AGN is shifted
towards the highest values: median $\log({\rm SFR}/M_\odot\,\hbox{yr}^{-1})=2.33$ with standard deviation $\sigma_{\log({\rm
SFR})}=0.90$. SFGs have a median $\log({\rm SFR}/M_\odot\,\hbox{yr}^{-1})=1.59$
with a standard deviation $\sigma_{\log({\rm SFR})}=0.81$. The median stellar masses of
all source populations are quite large: $\log(M_\star/M_\odot)= 11.05$, 11.17
and 10.89, for RL AGN, RQ AGN and SFGs respectively; the corresponding
standard deviations are 0.43, 0.59 and 0.51, respectively.

These median values vary with redshift. In the case of RL AGN the median $\log({\rm SFR}/M_\odot\,\hbox{yr}^{-1})$ increases from $-0.16$ at $z=0.5$ to 0.70 at $z=1$, to 1.67 at $z=2$; the corresponding dispersions are 0.63, 0.61 and 0.42, respectively. The relatively low dispersion at the highest redshift reflects the fact that at high-$z$ most RL AGN are hosted by star-forming galaxies while the fraction of passive hosts is large at low $z$.  Instead, the median $\log(M_\star/M_\odot)$ is quite stable; at the aforementioned redshifts it takes the values of 11.02, 11.09 and 11.00 with dispersions of 0.38, 0.33 and 0.44, respectively.

The increase with redshift of the mean SFR is somewhat less steep for RQ AGN and for SFGs, compared to RL AGN. At the same 3 redshifts, the median $\log({\rm SFR})$ is 0.95 (0.34), 1.53 (0.36) and 2.29 (0.34) for RQ AGN (standard deviations in parenthesis); for SFGs it is 1.02 (0.28), 1.58 (0.28) and 2.20 (0.27). At variance with RL AGN, also the median $\log(M_\star)$ of RQ AGN and SFGs increase with redshifts. For RQ AGN we have $\log(M_\star)=10.75$ (0.40), 10.95 (0.42) and 11.20 (0.42). For SFGs, $\log(M_\star)=10.68$ (0.32), 10.92 (0.30) and 11.12 (0.36). The specific SFR (sSFR) of SFGs increases from $\simeq 0.22\,\hbox{Gyr}^{-1}$ at $z=0.5$ to $\simeq 0.66\,\hbox{Gyr}^{-1}$ at $z=1$, to $\simeq 0.83\,\hbox{Gyr}^{-1}$ at $z=2$. The evolution of the sSFR with redshift is substantially shallower than found, in this redshift range, by several previous studies \citep[e.g.,][$\hbox{sSFR}\propto (1+z)^{p}$ with $p$ in the range 2.6--3]{Whitaker2014, Speagle2014, Johnston2015, Lehnert2015} but consistent with the results by \citet{Schreiber2015} and \citet[][for the sSFR derived from IR$+$UV data]{Bourne2017}, $p\simeq 1.5$--1.6.

The redshift distribution of each source sub-population is shown in
Fig.~\ref{fig:z_distributions}. The median redshift is $\sim 1.06$. The AGN
fraction of detected sources increases with increasing redshift. \citet{Duncan2020} mention the possibility of selection effects in the spectroscopic sample used for photo-$z$ training. They also caution that the limited number of spectroscopic redshifts at $z>1$ results in a deterioration of the quality of photometric redshift estimates for host-dominated sources in the range $1<z<1.5$, and the quality becomes substantially worse for $z>1.5$. 

The origin of the dip at $z\sim 1.3$ can be largely ascribed to aliasing effects in the photo-$z$ distribution. The photo-$z$'s were fixed at the median
of the first photometric redshift solution, and for reasons associated with the filter wavelength distribution, this seems to disfavour redshifts $z\sim 1.3$ for both SFGs and RL AGN and instead alias them to slightly higher or lower redshifts.
We note, incidentally, that the very small fraction of spectroscopic redshifts at $z>1$ implies that our redshift distribution is not biased by the ``redshift desert'', i.e. by the lack of strong spectroscopic features detectable from the ground for galaxies in the redshift range $z\sim 1.4$--2.5 \citep[e.g.,][]{Steidel2004}.

On the other hand, a feature in the redshift distribution of star-forming galaxies at $z\sim 1.5$ is expected since the dominant population of star-forming galaxies changes around this redshift: below $z\sim 1.5$ star-forming galaxies are mostly late-type while at higher redshifts we enter the era when early-type galaxies formed the bulk of their stars \citep{Cai2013}. Hints of a dip at $z\sim 1.5$ can be discerned in the redshift distribution for $S_{1.4\,\rm GHz}>50\,\mu$Jy (not far from the effective depth of the LoTSS survey) predicted by \citet{Mancuso2015b}. According to the model, the dip is caused by the fact that, at $z\simeq 1.5$, the rapid drop of the star-formation activity of proto-spheroidal galaxies, which formed most of their stars at higher redshifts, is only partly compensated by normal late-type and starburst galaxies, which dominate the star-formation at lower redshifts. Since the radio  luminosity functions of proto-spheroidal and late-type/starburst galaxies are substantially different, the presence/absence of the dip depends on the luminosity corresponding to the flux density limit of the survey at $z\simeq 1.5$; in other words, the presence or absence of the dip depends on the flux density limit.

\begin{figure*}
\begin{center}
\includegraphics[width=0.49\textwidth]{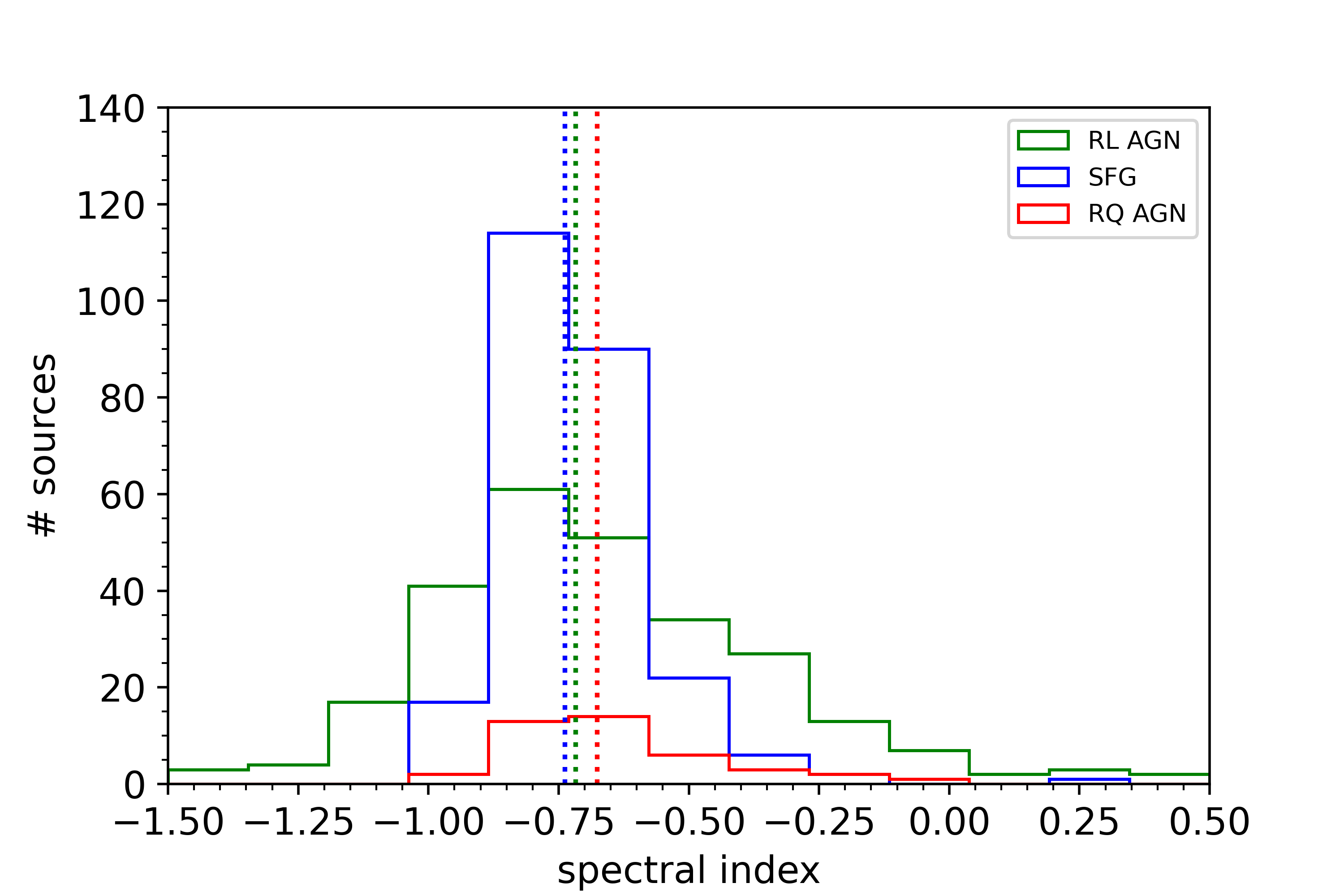}
\includegraphics[width=0.49\textwidth]{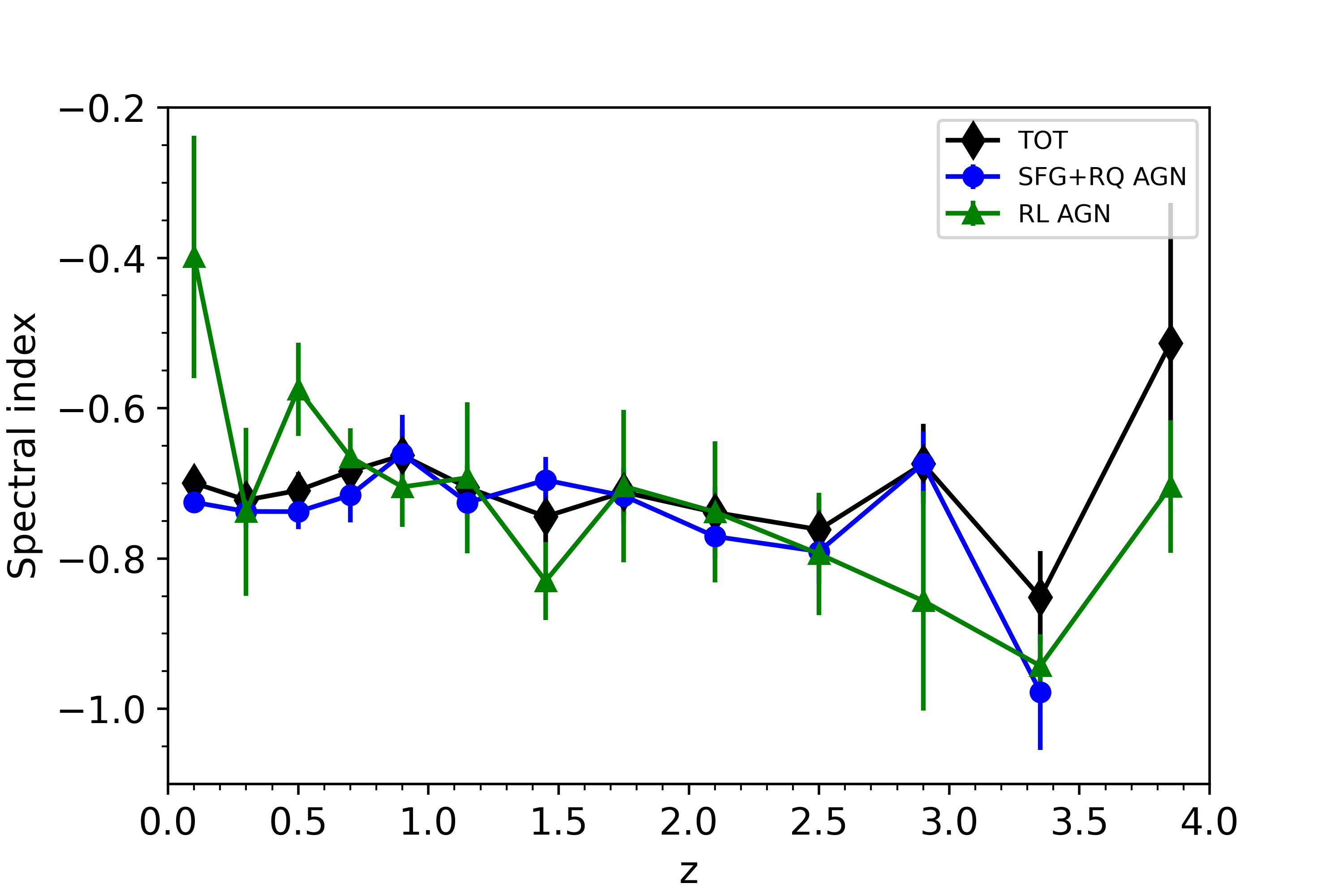}
\caption{Left panel: spectral index distributions of the 3 source populations.
The median spectral indices (vertical lines) are: $-0.72$ for RL AGN,
$-0.74$ for SFGs and $-0.68$ for RQ AGN. Right panel: redshift dependence of the
median spectral indices of sources classified as RL AGN (filled green triangles) and  as non-RL AGN (i.e., as SFGs or RQ AGN; filled blue circles) based on both WSRT and LOFAR data. The filled black diamonds (``TOT'') represent the median values for the full set of WSRT sources with LOFAR counterparts, including those with different classifications and the unclassified ones.    Error bars are the interquartile ranges divided by the
square root of the number of objects in the bin. A slight trend towards a steepening of the
spectral index with increasing $z$ is visible, especially in the case of RL AGN, but its statistical significance is low (see text).
}
 \label{fig:histo_si}
  \end{center}
\end{figure*}

\section{Results}\label{sec:results}

\subsection{Comparison with an independent 1.4\,GHz based source classification}\label{subsec:spectral_indices}

\citet{Bonato2020} studied a sample of 1173 sources brighter than $S_{1.4\,\rm
GHz} = 120\,\mu$Jy, i.e. with a signal-to-noise ratio $\ge 10$, in the central $\simeq 1.4\,\hbox{deg}^2$ of the $\simeq 6.6\,\hbox{deg}^2$ area covered by the WSRT survey in LH field by \citet{Prandoni2018}; the synthesized beam is
of $11''\times 9''$ with position angle $\hbox{PA}=0^\circ$. Thus this sample covers an area about a factor of seven smaller than that covered by the present LOFAR sample. SFRs were taken, when available, from the \textit{Herschel} Extragalactic Legacy Project (HELP) catalogue \citep{Shirley2019}; otherwise the $24\,\mu$m flux density was used as a SFR indicator \citep{Battisti2015}.

The HELP SFRs were obtained via SED fitting with CIGALE \citep{Malek2018}. HELP collates, heals and homogenises multi-frequency data over the $1270\,\hbox{deg} ^2$ covered by the \textit{Herschel} SPIRE extragalactic surveys. The surveys covering each \textit{Herschel} field are highly variable in number, wavebands and depth.
For the WSRT LH field, HELP contains optical photometry in the $g$, $r$, $i$, $z$, and $y$ bands and near-IR photometry in the $K$ band and in \textit{Spitzer} IRAC bands, in addition to \textit{Herschel} data. CIGALE was run for all HELP galaxies with at least two optical and at least two near-IR detections \citep{Shirley2021}. The code estimates the galaxy properties by SED fitting. Model SEDs are computed dealing self-consistently with stellar emission and its reprocessing by dust.

For the comparison with the classification based on the LOFAR flux densities we have redone the classification after rescaling the WSRT flux densities by a factor of 0.84 in order to bring them to statistical agreement with the multi-frequency radio measurements in the same field (see paper II; their Fig.\,D.1). This has led to some differences with the original classification by \citet{Bonato2020}. Apart from that, the classification was done using the same criteria. If an estimate of the SFR was available, sources were classified as RL AGN if their $L_{1.4\rm GHz}/\hbox{SFR}$ ratios exceeded the redshift-dependent threshold derived by \citet{Delvecchio2017}. Otherwise, the criterion by \citet{Magliocchetti2017}, based on radio luminosity, was adopted. Non-RL AGN sources were classified as either SFGs or RQ AGN using the diagnostics, based on near/mid-IR colours, by \citet{Messias2012} or by \citet{Donley2012}, depending on the available data. The classification was tested by exploiting the X-ray data from the XMM-Newton survey by \citet{Brunner2008}, which extends over $\simeq 10\%$ of the field covered by the \citet{Bonato2020} sample.

Most of the WSRT sources ($1004/1173 \simeq 86\%$) have a LOFAR counterpart within 3.5\,arcsec, corresponding to 3 times the quadratic sum of WSRT \citep{Prandoni2018} and LOFAR \citep{Kondapally2020} rms positional uncertainties, $\simeq 1''$ and $\simeq 0.6''$, respectively. The WSRT astrometric error refers to the $5\,\sigma$ flux density limit; the LOFAR error is a conservative empirical estimate. The surface densities of the WSRT sample studied by \citet{Bonato2020} and of the present LOFAR sample are $\simeq 6.46\times 10^{-5}\,\hbox{arcsec}^{-2}$ and $\simeq 2.33\times 10^{-4}\,\hbox{arcsec}^{-2}$, respectively. In the absence of further information, the expected number of chance associations within 3.5\,arcsec would be $\simeq 10.5$ ($\simeq 0.9\%$). However, in making the WSRT/LOFAR associations we have taken into account also the optical/near-IR identifications which have substantially better positional accuracy. In this way the number of chance associations becomes negligible.


It is interesting to compare the classifications by \citet{Bonato2020} with those by paper V. The WSRT--based classification of the 1004 common sources yielded  392 RL AGN, 306 SFGs, 131 RQ AGN and 175 unclassified. The LOFAR--based classification yielded 379 RL AGN, 493 SFGs, 93 RQ AGN and 39 unclassified.  The two classifications agree for 86\% of sources classified as non-RL AGN, demonstrating a remarkable stability against variations of the classification criteria. The situation is trickier in the case of RL AGN. Only $\simeq 69\%$ of sources classified in this way by \citet{Bonato2020} have the same classification in paper V; about 26\% of sources become  SFGs or RQ AGN while the remaining $\simeq 5\%$ are unclassified.

Some classification discrepancies are expected because of differences in the classification criteria (50 out of the 104 LOFAR-unconfirmed WSRT RL AGN were classified using the  criterion by \citealt{Magliocchetti2017}), in estimates of the SFRs (also due to different input photometry) and because of the different radio excess threshold. However, since the adopted thresholds are quite similar, the most likely culprit at least for a fraction of sources is a difference in the radio luminosity to SFR ratio between 1.4\,GHz and 150\,MHz, compared to the expectation for the average spectral index. We have therefore investigated the spectral index distributions. They are shown in the left-hand panel of Fig.~\ref{fig:histo_si}. The median spectral index for the full sample is $-0.70$. The median spectral indices of the
various populations are close to each other: $-0.72$, $-0.68$ and $-0.74$,
respectively, for sources classified as RL AGN, RQ AGN and SFGs by both
\citet{Bonato2020} and \citet{Best2020}. The median spectral index of non-RL AGN (i.e., SFGs plus RQ AGN) is $-0.73$.

Sources classified as RL AGN by \citet{Bonato2020} but not by \citet{Best2020} have a substantially flatter
median spectral index ($-0.55$, corresponding to a difference of about 0.15 dex in 150 MHz radio luminosity), implying lower $L_{150\,\rm MHz}$/SFR ratios than expected based on the median spectral index of the full sample. They can thus fall below the threshold for classification as RL AGN. In fact we find that in most cases the origin of the different classification can be traced back to this effect. A similar conclusion applies to the 47 WSRT RL AGN without a LOFAR counterpart\footnote{There are 169  WSRT sources without a LOFAR counterpart; 47 of them are classified as RL AGN, 23 as SFGs, 5 as RQ AGN and 94 are unclassified.}.

It should be noted that the impact of this problem on our analysis is modest. In fact RL AGN comprise a minor fraction of our sources; they are only 22\% of non-radio excess sources. Thus even in the worst case the number of SFGs plus RQ AGN would have been overestimated by $0.27\times 0.22 \simeq 6\%$. However, the fact that these sources fall below the threshold for the classification as RL AGN at 150\,MHz means that, at this frequency, a significant fraction, probably the majority, of the radio emission is due to star formation. Hence the comparison between the abundances of different source populations measured at different frequencies requires some caution.
In the following we use the classification by \citet{Best2020}.

The right-hand panel of Fig.~\ref{fig:histo_si} shows an indication of a flattening of the median spectral index of RL AGN at the lowest redshifts. The indication is only marginally significant (at the $\simeq 95\%$ confidence level, based on Pearson's correlation coefficient) for our sample, but a similarly flat spectral index of low-$z$ AGN was reported by \citet[][their Fig.\,10]{CalistroRivera2017} for a brighter sample, although they found a steepening of the AGN spectral index between 150 and 325\,MHz. Apart from that, the figure shows a slight trend towards a steepening of the AGN spectral index with increasing $z$. This kind of trend has long been known, although its origin is not well understood \citep[see][for a review]{MileyDeBreuck2008}. The statistical significance of the redshift dependence is however low (the probability of an occurrence by chance is 0.01, corresponding to $2.57\,\sigma$), consistent with the earlier conclusions by \citet{CalistroRivera2017}.




\begin{figure*}
\begin{center}
\includegraphics[width=0.98\textwidth]{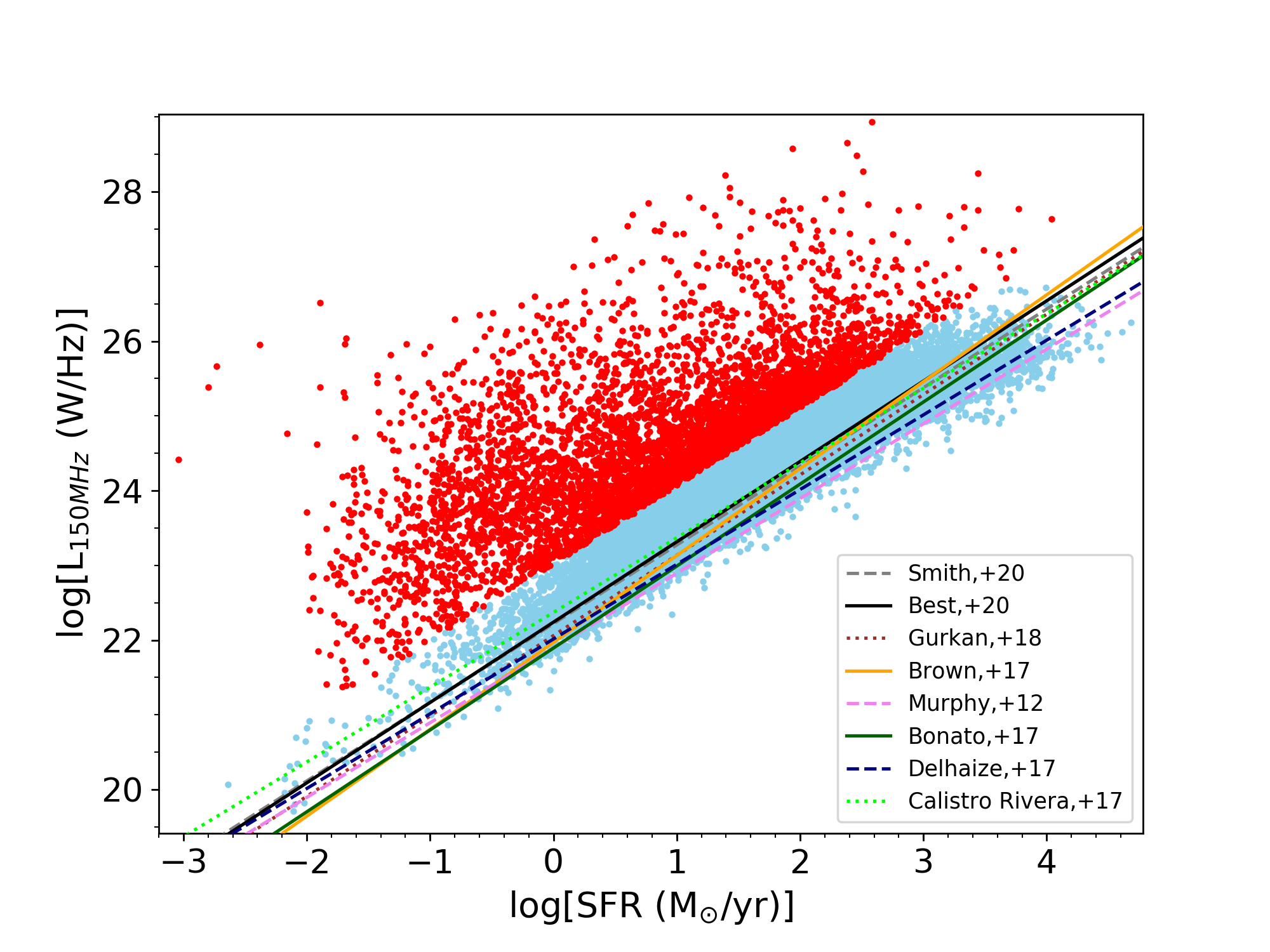}
\caption{Radio luminosity at 150\,MHz versus SFR. Red and cyan points represent, respectively, RL AGN
and non-RL sources (i.e., SFGs plus RQ AGN). The black solid line shows the ridge line in the
radio--SFR relation used by \citet{Best2020} to define the radio excess threshold (0.7 dex above the ridge line)
separating RL AGN from SFGs plus RQ AGN.
Also shown, for comparison, are the relations by \citet[grey dashed line]{Smith2020},
by \citet[][brown dotted line]{Gurkan2018}, \citet[][non-linear model; dark green solid line]{Bonato2017},
\citet[][dark blue dashed line]{Delhaize2017}, \citet[][orange solid line]{Brown2017},
\citet[][light green dotted line]{CalistroRivera2017} and \citet[][pink dashed line]{Murphy2012}.}
 \label{fig:SFR_Lradio}
  \end{center}
\end{figure*}

\begin{figure}
\begin{center}
\includegraphics[width=0.49\textwidth]{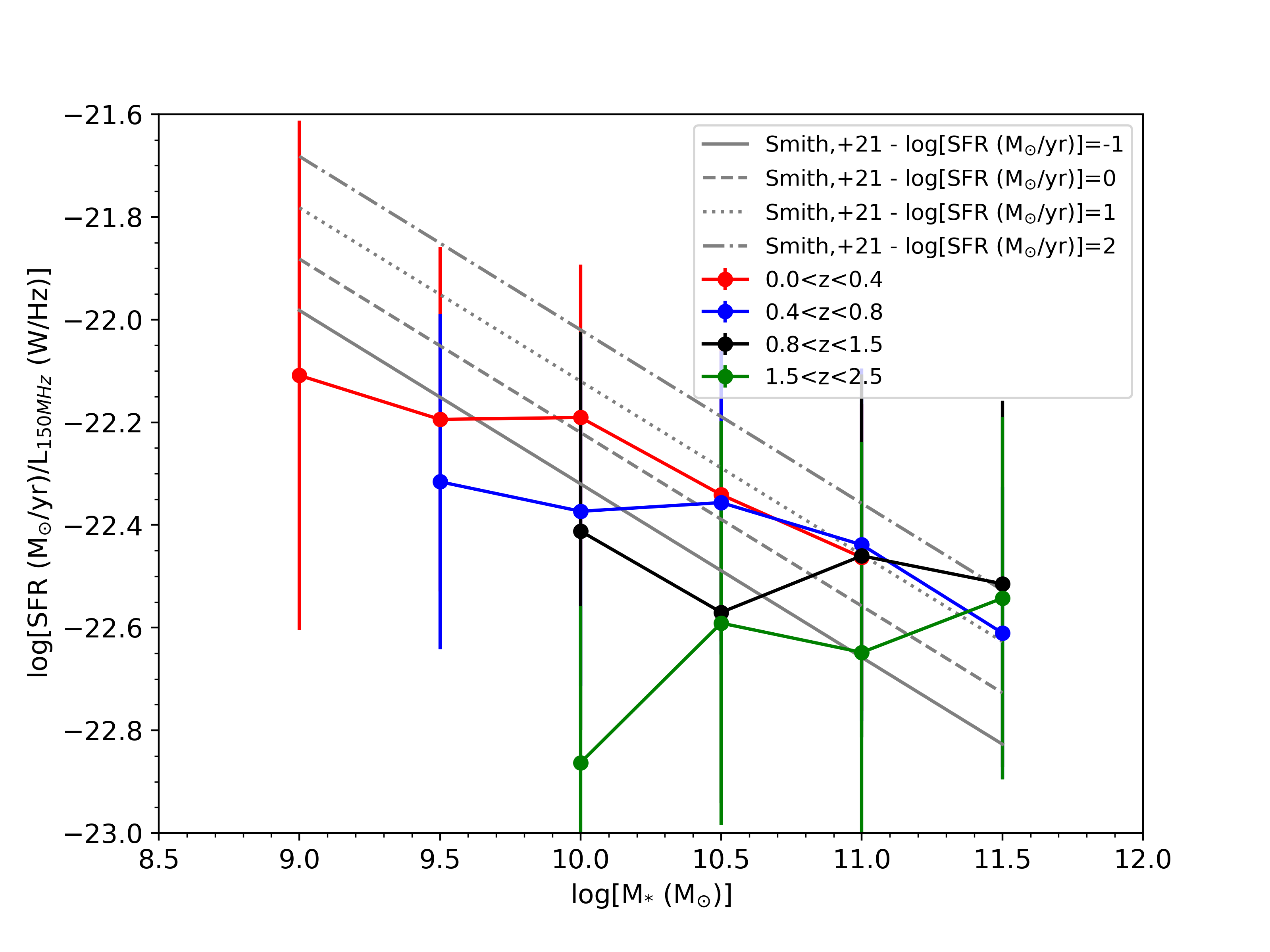}
\caption{Ratio of the radio luminosity at 150\,MHz to the SFR as a function of $M_\star$ of SFGs plus RQ AGN
for four redshift bins. The points are the modes of the distributions of $\log(\hbox{SFR}/L_{150\,\rm MHz})$. The error bars are the interquartile errors. The straight lines show, for comparison, the relations derived by \citet{Smith2020} for 4 values of the SFR. Although these relations are steeper than ours, the agreement is good for SFGs with typical properties, i.e. where we have more data, both globally and for each redshift bin (see text).}
 \label{fig:SFR_Lradio_Mstar}
  \end{center}
\end{figure}

\begin{figure}
\begin{center}
\includegraphics[width=0.49\textwidth]{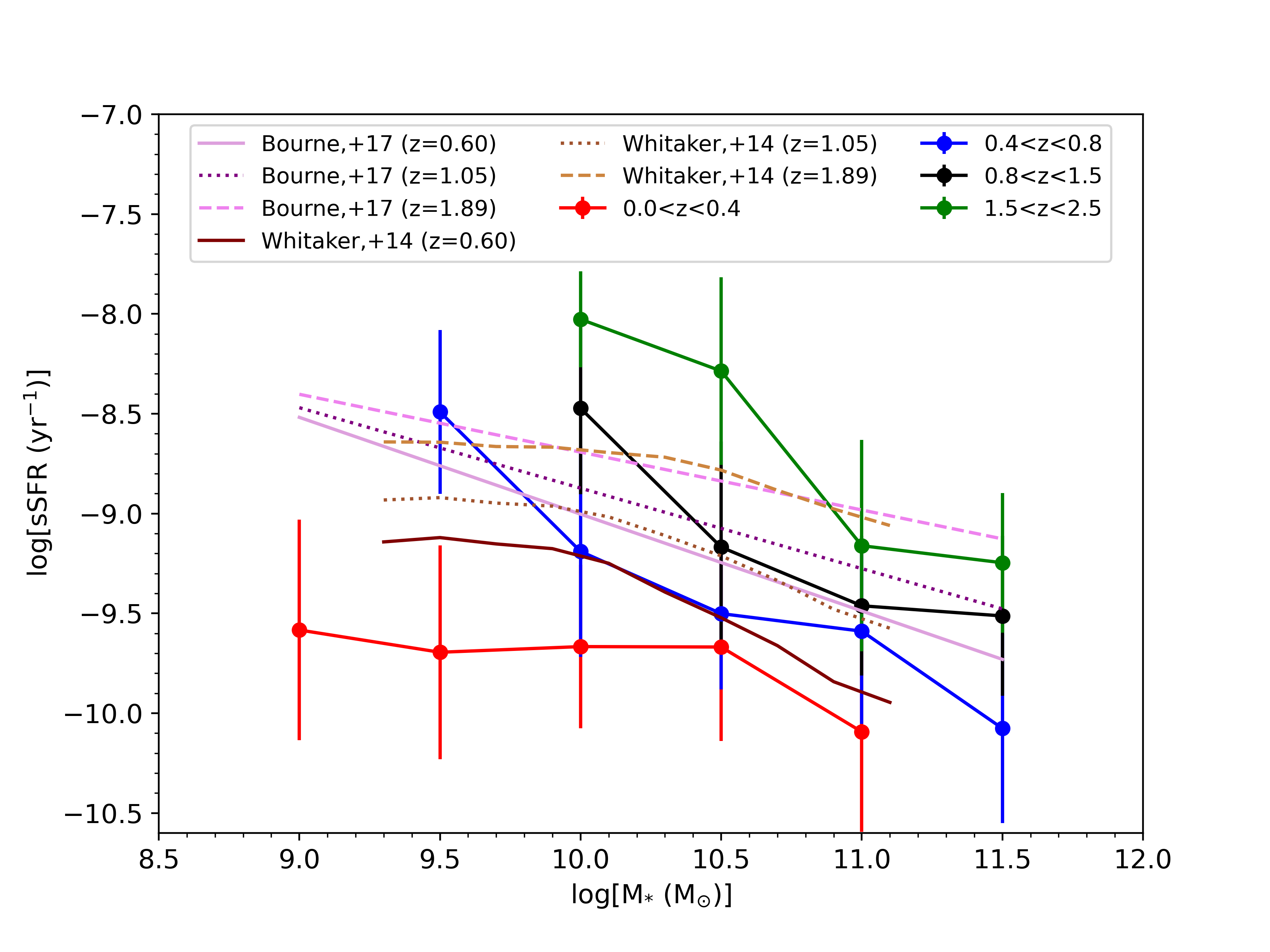}
\caption{Specific SFR of our SFGs plus RQ AGN as a function of $M_\star$ for 4 redshift bins. The lines show the analytic models by \citet{Whitaker2014} and \citet{Bourne2017}, fitting their data, computed at the median redshifts of our bins ($z=0.60$, 1.05 and 1.89). Our lowest redshift bin is outside the range covered by the other studies.  }
 \label{fig:sSFR}
  \end{center}
\end{figure}

\subsection{Radio Luminosity versus SFR}\label{subsec:SFR_Lradio_relation}

Figure\,\ref{fig:SFR_Lradio} displays  the 150\,MHz radio luminosity versus SFR
for our sample. The cyan points identify the ``SFG region'' (including RQ AGN). Some sources that were classified as RL AGN on the basis of clearly extended radio emission (i.e. jets) turned out to be located within the ``SFG region''. They could be radio-loud AGN where the radio excess just does not quite make the 0.7 dex threshold or RQ AGN, as they can also have jets, and have been reclassified as such. These objects are a tiny fraction of sources and their reclassification does not significantly affect our results.

The black solid line is the ridge-line in the radio--SFR relation used by \citet{Best2020} to define the radio excess threshold separating RL AGN from SFGs and RQ AGN. It is given by $\log({\rm SFR}) = 0.93 (\log L_{150\,\rm MHz} - 22.24)$ with SFR in
$M_\odot\,\hbox{yr}^{-1}$ and $L_{150\,\rm MHz}$ in ${\rm W\,Hz^{-1}}$. The use of the ridge line mitigates the effect of the radio selection, which biases the mean or median values towards brighter radio sources. As mentioned above the upper bound of the ``SFG region'' is set at $L_{150\,\rm MHz}$ 0.7 dex above the ridge line.

The figure also shows, for comparison, the stellar mass-independent relations presented by \citet{Smith2020}, \citet{Gurkan2018}, \citet{Brown2017}, \citet{Delhaize2017},
\citet{CalistroRivera2017}, \citet[][non-linear model]{Bonato2017} and
\citet{Murphy2012}. The Smith et al. mass-independent relation was derived for a near-infrared selected sample of $z<1$ galaxies, using the LOFAR survey data over the ELAIS-N1 field. The SFRs were obtained from MAGPHYS alone, rather
than from the combination of SED fitting methods used by \citet{Best2020}.
To avoid the bias related to the radio selection, \citet{Smith2020} used all estimates of $L_{150\,\rm MHz}$, irrespective of their statistical significance, since they showed that censoring the sample could affect the derived SFR/$L_{150\,\rm MHz}$ relation. The resultant relation is in fairly close agreement with the ridge line of \citet{Best2020}.

The relations presented by \citet{Gurkan2018} and by
\citet{CalistroRivera2017} are based on shallower LOFAR surveys. The former authors exploited a sample selected from
the seventh data release of the Sloan Digital Sky Survey \citep[SDSS
DR7;][]{Abazajian2009} catalogue over the Herschel Astrophysical Terahertz
Large Area Survey \citep[H-ATLAS;][]{Eales2010} North Galactic Pole (NGP) field
\citep{Maddox2018} which overlaps with the LoTSS DR1 (\citealt{Shimwell2017}). SFRs and stellar masses were computed using MAGPHYS. In
Fig.\,\ref{fig:SFR_Lradio} we show their single power-law fit, although they
also derive a stellar-mass dependent relation.

\citet{CalistroRivera2017} investigated the IR-radio correlation at both 150\,MHz and 1.4\,GHz of radio
selected SFGs over the LOFAR Bo\"otes field \citep{Williams2016}. The source
classification and the determination of their physical parameters were
performed using the SED-fitting algorithm AGNfitter. The IR-radio correlation
was found to be redshift-dependent. The line shown in
Fig.\,\ref{fig:SFR_Lradio} refers to the median redshift of our sample
($z=1.06$).

The study by \citet{Delhaize2017} relies on the high-sensitivity VLA
observations at 3\,GHz in the COSMOS field. Luminosities at 1.4\,GHz were
computed using the individual spectral indices when available or $\alpha=-0.7$
otherwise. SFRs were derived from total IR luminosities using the conversion by
\citet{Kennicutt1998} assuming a \citet{Chabrier2003} initial mass function
(IMF). We have extrapolated their infrared-to-radio luminosity ratio at
1.4\,GHz to 150\,MHz by setting $\alpha=-0.7$. A slight decrease of such ratio
with increasing redshift was reported by \citet{Delhaize2017}. Again, the line shown in Fig.\,\ref{fig:SFR_Lradio}
refers to $z=1.06$.

\citet{Murphy2011} calibrated the 1.4\,GHz radio luminosity versus SFR relation
using Starburst99 \citep{Leitherer1999} for a \citet{Kroupa2001} IMF in
combination with some empirical recipes.

\citet{Bonato2017} computed the local radio luminosity function at 1.4\,GHz by
convolving the observationally determined local SFR function
\citep{Mancuso2015b} with a power-law dependence of $L_{1.4}$ on the SFR. The
parameters of the $L_{1.4}$--SFR relation were derived by fitting the local
1.4\,GHz luminosity function of SFGs \citep{MauchSadler2007}. The Murphy et al.
and Bonato et al. relations were extrapolated to 150\,MHz using the thermal and
non-thermal spectral indices quoted in the respective papers.

As illustrated by Fig.\,\ref{fig:SFR_Lradio}, there are differences among the
relations, both in slope and in normalization. The variations of the
$L_{150\,\rm MHz}/\hbox{SFR}$ ratios can reach a factor $> 2$ (cf. Sect.~\ref{sec:introduction}).

It can be noted that, at low SFRs, the relation from paper V
(black solid line in Fig.\,\ref{fig:SFR_Lradio}) is lying below the average
value of detected sources. This is due to having adopted the ridge line instead
of the mean to minimize the effect of radio selection.

Figure~\ref{fig:SFR_Lradio_Mstar} shows the mode of the distribution of $\hbox{SFR}/\log(L_{150\,\rm MHz})$  ratios of SFGs plus RQ AGN (data points) as a function of stellar mass for four redshift bins. The error bars are simply the interquartile ranges within each bin. These are intended to give an indication of the range of values: a proper error estimate would involve taking into account the probability distributions of the photometric redshifts, and uncertainties on consensus stellar masses and SFRs, which is beyond the scope of this paper. Furthermore, the measured values are potentially affected by systematic factors, such as the radio flux density limits.

We see a slight trend towards a lower $\hbox{SFR}/\log(L_{150\,\rm MHz})$ ratio with increasing $M_\star$ at $z\simlt 1$. This trend is not seen at higher redshifts, and may even be reversed in the bin $1.5<z<2.5$. The trend is substantially weaker than those reported by \citet{Smith2020} for $z<1$, and by \citet{Delvecchio2020} for $0.1< z < 4.5$. These authors find, respectively, a decrease of $0.33\pm 0.04$ and of $0.234\pm 0.017$ per decade of $M_\star$ above $\log(M_\star/M_\odot)=10$ while our data indicate a decrease $\lesssim 0.1$.

These differences reflect the different primary selection. Our radio selection favours higher radio luminosities for a given SFR, hence lower $\log(\hbox{SFR}/L_{150\,\rm MHz})$. This selection bias is increasingly important for lower mass galaxies, which are expected to have lower SFRs and hence lower radio luminosities; this leads to an increasing fraction falling below the radio flux density limit and hence a potential bias in the measured value of $\log(\hbox{SFR}/L_{150\,\rm MHz})$ as only the objects with the brightest $L_{150\,\rm MHz}$ for a given SFR are detected.  It may well account for the flatter slope of the $\log(\hbox{SFR}/L_{150\,\rm MHz})$ versus $M_\star$ relation compared to those found by \citet{Smith2020} and \citet{Delvecchio2020} who started from near-infrared and from $M_\star$--selected samples, respectively.
Note that, even though the two latter selections are similar in that they favour higher $M_\star/\hbox{SFR}$ ratios, they still lead to significantly different relation.  This is another manifestation of the importance of selection effects. However, we are in good agreement with the results by \citet{Smith2020} where we have more data, i.e. for SFGs with typical properties in each redshift bin. This can be seen considering the peaks of our distributions of SFRs and of stellar masses: they occur at SFRs of 1.1, 9.2, 29.4, $83.1\,M_\odot\,\hbox{yr}^{-1}$ and $\log(M_\star/M_\odot)=10.49$, 10.80, 11.04, 11.12 for $0<z<0.4$,  $0.4<z<0.8$, $0.8<z<1.5$, $1.5<z<2.5$, respectively. The peaks of the global distributions occur at $\log(M_\star/M_\odot)\simeq 11$, $z\sim 1$, $\log(\hbox{SFR}/M_\odot\,\hbox{yr}^{-1})\simeq 1.6$.

Figure\,\ref{fig:SFR_Lradio_Mstar} also shows an indication of a decrease of the
$\log(\hbox{SFR}/L_{150\,\rm MHz})$ ratio with increasing redshift, except for the largest stellar masses. In terms of $q_{\rm TIR}\propto (1+z)^\delta$  we find $\delta \simeq -0.12$  for $\log(M_\star/M_\odot)\simeq 10.5$, comparable to the values reported by \citet[][$\delta=-0.19 \pm
0.01$]{Delhaize2017}, by \citet[][$\delta=-0.12 \pm 0.04$]{Magnelli2015} and by
\citet[][$\delta=-0.22 \pm 0.05$, at 150\,MHz]{CalistroRivera2017}. The evidence for a redshift evolution progressively weakens with increasing $M_\star$ and becomes insignificant for $\log(M_\star/M_\odot)\simgt 11$. We caution, however, that the higher redshifts are more likely to suffer from the biases discussed above.

Our results are consistent with the conclusion by \citet{Smith2020} and  \citet{Delvecchio2020} that the cosmic evolution
of the $\hbox{SFR}/L_{150\,\rm MHz}$ ratio is mostly driven by its decrease
with increasing $M_\star$, coupled with the fact that the effective $M_\star$
of radio-selected SFGs increases with redshift \citep[see the upper panel of Fig.~3 of][]{Smith2020}.

Figure~\ref{fig:sSFR} presents a different view of our data by showing them in terms of the specific SFR (sSFR) as a function of $M_\star$. There is a clear decrease of the sSFR with increasing $M_\star$ except perhaps in the lowest redshift bin. A similar trend was reported by \citet{Bourne2017} who used the deepest 450-- and 850--$\mu$m imaging from SCUBA-2 Cosmology Legacy Survey. Our results are consistent with  theirs, within the errors of both estimates, in the overlapping redshift range ($0.5< z < 2$), although our data hint at an excess sSFR of the lowest stellar masses. If real, the excess may be due to a higher efficiency of the radio selection at detecting early phases of galaxy evolution, when the stellar masses were still relatively low. Our data do not show the flattening of the slope of the $\log(\hbox{sSFR})$ vs. $\log(M_\star)$ relation at low $M_\star$ reported by \citet{Whitaker2014} for their near-IR selected sample. This can be another indication that the optical/near-IR selection underestimates the SFR of dust-enshrouded galaxies in their early evolutionary phases.

As pointed out above, the redshift evolution of the amplitude of the  $\log(\hbox{sSFR})$ vs. $\log(M_\star)$ relation implied by our data is consistent with the results by \citet{Bourne2017} but is substantially weaker than found by \citet{Whitaker2014} and by other surveys (cf. sub-sect.~\ref{sect:classification}).

\begin{figure}
\begin{center}
\includegraphics[width=0.49\textwidth]{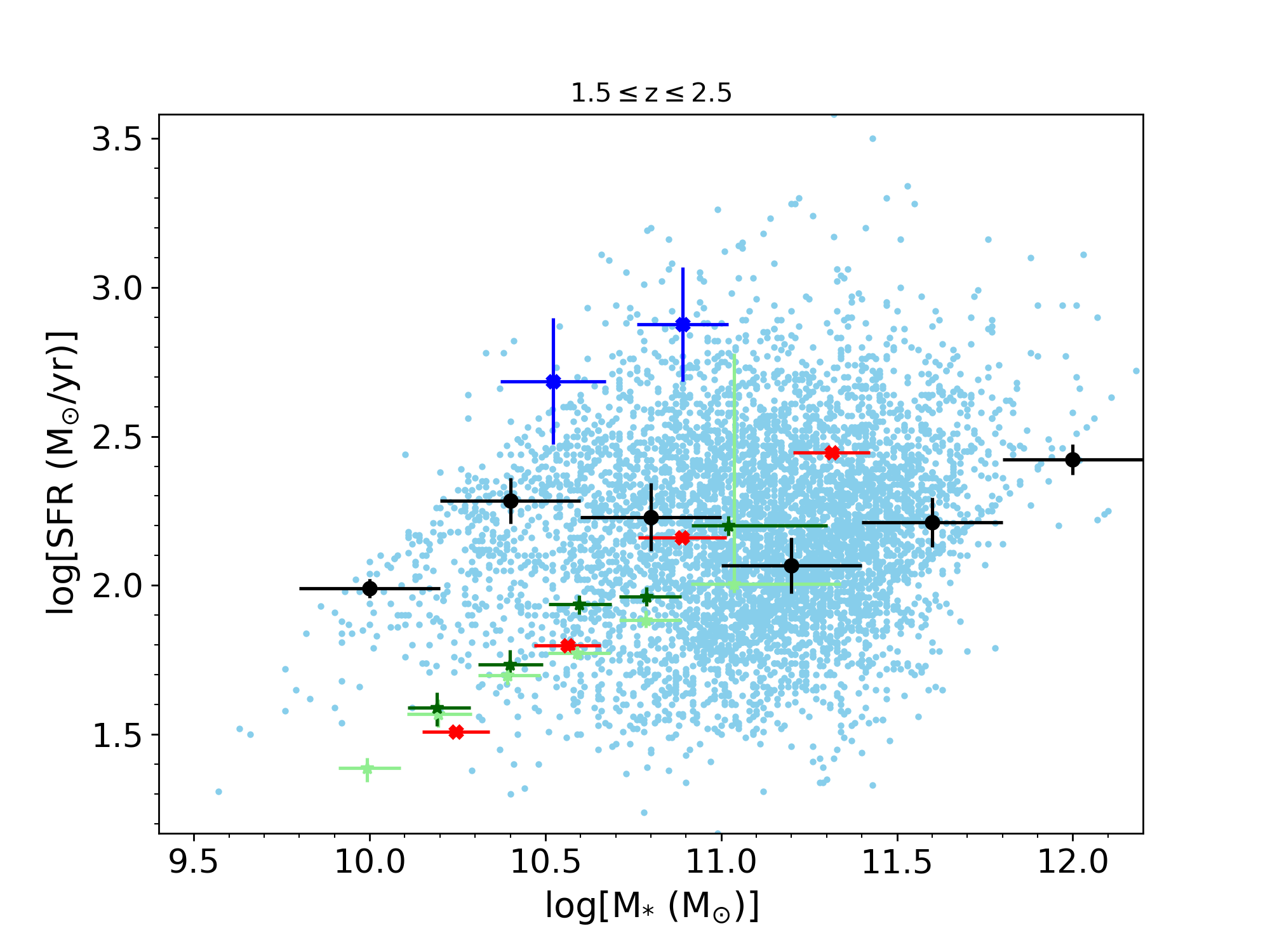}
\caption{Distribution in the SFR--$M_\star$ plane of our SFGs and RQ AGN in
the $1.5 \leq z \leq 2.5$ redshift interval (cyan points). The black filled circles with error bars show the modes of the SFR distributions averaged over the stellar mass bins represented by horizontal bars. The numbers of sources in each bin, from low to high $M_\star$, are  93, 511, 1511, 2491, 1001 and 50, respectively. The vertical bars show the dispersions. The red symbols represent the average values for the \citet{Rodighiero2015}
K-band selected sample of star-forming galaxies.  The upper blue points show the average SFRs
for the \citet{Rodighiero2015} sample of starburst galaxies in their two mass bins. The green asterisks represent the \citet{Leslie2020} median values for their SFG sample (light green for their 1.5$-$2 redshift range, dark green for 2$-$2.5).}
 \label{fig:SFR_Mstar}
  \end{center}
\end{figure}

\subsection{SFR versus stellar mass}\label{subsec:SFR_Mstar_relation}

The correlation between SFR and $M_\star$ for SFGs, known as the ``main
sequence'' (MS), has been extensively discussed in the literature \citep[e.g.,][and
references therein]{Noeske2007, Elbaz2007, Rodighiero2014, Schreiber2017,
Santini2017, Bera2018, Gu2019, Leslie2020}. The shape of this correlation was found to vary depending
on the sample selection. Figure~\ref{fig:SFR_Mstar} shows the distribution in the $\log(\hbox{SFR})-\log(M_\star)$ plane of
our SFGs in the 1.5$-$2.5 redshift range studied by \citet{Rodighiero2015}, a reference paper for the definition of the main sequence. The correlation between
the two quantities is weak, as also found for sub-mm selected galaxies
\citep[e.g.,][]{Michalowski2012, Koprowski2016}.

Our radio-selected sample has an average $\hbox{SFR}/M_\star$ ratio similar to
that of the K-band selected sample of SFGs investigated by \citet{Rodighiero2015}
for $\log(M_\star/M_\odot)\sim 11$. At lower stellar masses our objects have
substantially higher ratios, as expected since our selection favours objects
with high SFRs while the optical selection favours lower $\hbox{SFR}/M_\star$ ratios. Above $\log(M_\star/M_\odot)\sim 11$ our average $\hbox{SFR}/M_\star$
ratios are below the extrapolation of the \citet{Rodighiero2015} relation, but
consistent with, or slightly larger than the mean ratios reported by
\citet{Leslie2020} for the VLA-COSMOS 3\,GHz Large Project sample at $z\lesssim
2.5$, where the overwhelming majority of our SFGs reside. This is in agreement
with the flattening/bending of the $\hbox{SFR}$--$M_\star$ relation at large
stellar masses reported by several authors \citep[e.g.,][]{Schreiber2015,
Tomczak2016} and investigated in great detail by \citet{Leslie2020}.

The distribution of SFRs of galaxies in our sample reaches substantially higher values than the deep K-band selected sample by \citet{Rodighiero2015}, seamlessly extending over the region of ``starburst'' galaxies which do not appear as a distinct group. In other words, our radio selection does not underscore any clear correlation between SFR and $M_\star$. Such correlation arises when star-forming galaxies have grown a sufficiently large stellar mass to allow an optical/near-infrared selection. The radio and FIR/sub-mm selections are insensitive to the stellar mass and therefore yield samples covering a very wide range of $M_\star$/SFR ratios.

Our radio selection provides a view of the distribution of galaxies in the $\hbox{SFR}$--$M_\star$ complementary to that of optical/near-IR selection. The latter emphasizes $M_\star$ while the radio (and the far-IR/sub-mm) selection hinges upon the SFR. To be more specific, at high stellar masses, the radio selection picks out objects that are beginning to quench (below the main sequence) and yet still have sufficient SFR to be detected; these are classified as SFGs in the current analysis because they have no AGN activity, but in typical optical selections they are excluded from main sequence analyses as ``quenched'' galaxies. At lower masses, the radio observations do not have the sensitivity to probe the typical objects seen on the mass-selected main sequence; instead the radio sample is dominated by objects above the main sequence (starbursting, or just randomly high). This gives a different picture.

Both views must be taken into account in order to properly understand the star-formation history of galaxies. For example, the distinction between ``main-sequence'' and ``off-sequence'' galaxies, frequently found in the literature \citep[e.g.,][]{Rodighiero2011, Rodighiero2015, Silverman2015, Schreiber2015}, tends to be interpreted as indicative of two different modes of star-formation: a secular star-formation mode of galaxies within the MS and a stochastic star-formation mode with strongly enhanced star-formation efficiency, perhaps triggered by major mergers for ``off-sequence'' galaxies \citep[e.g.,][]{Elbaz2018}.

However, the fact that the radio selection does not show any clear sequence suggests a higher uniformity of the star formation history of galaxies, with the MS corresponding to a long-lived phase of galaxy evolution, with a well developed stellar population, most easily picked up by the optical/near-IR selection. Instead, the radio (and far-IR/sub-mm) selection sees the full history of star-formation, from the initial phases when the stellar masses were very low (i.e. the $\hbox{SFR}/M_\star$ ratios were well above the main sequence) to later phases when these ratios take on MS values.

\begin{figure*}
\begin{center}
\includegraphics[width=0.49\textwidth]{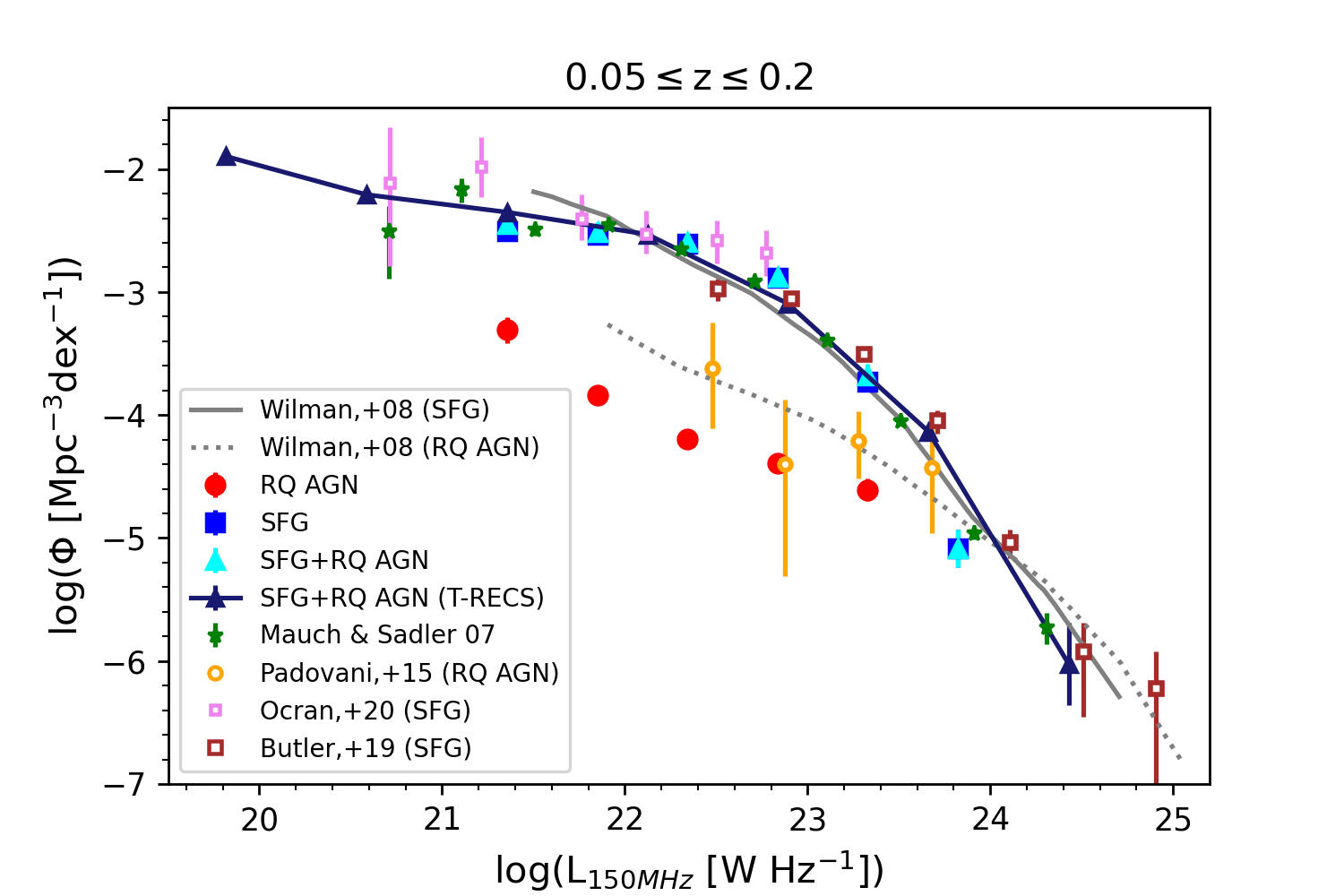}
\includegraphics[width=0.49\textwidth]{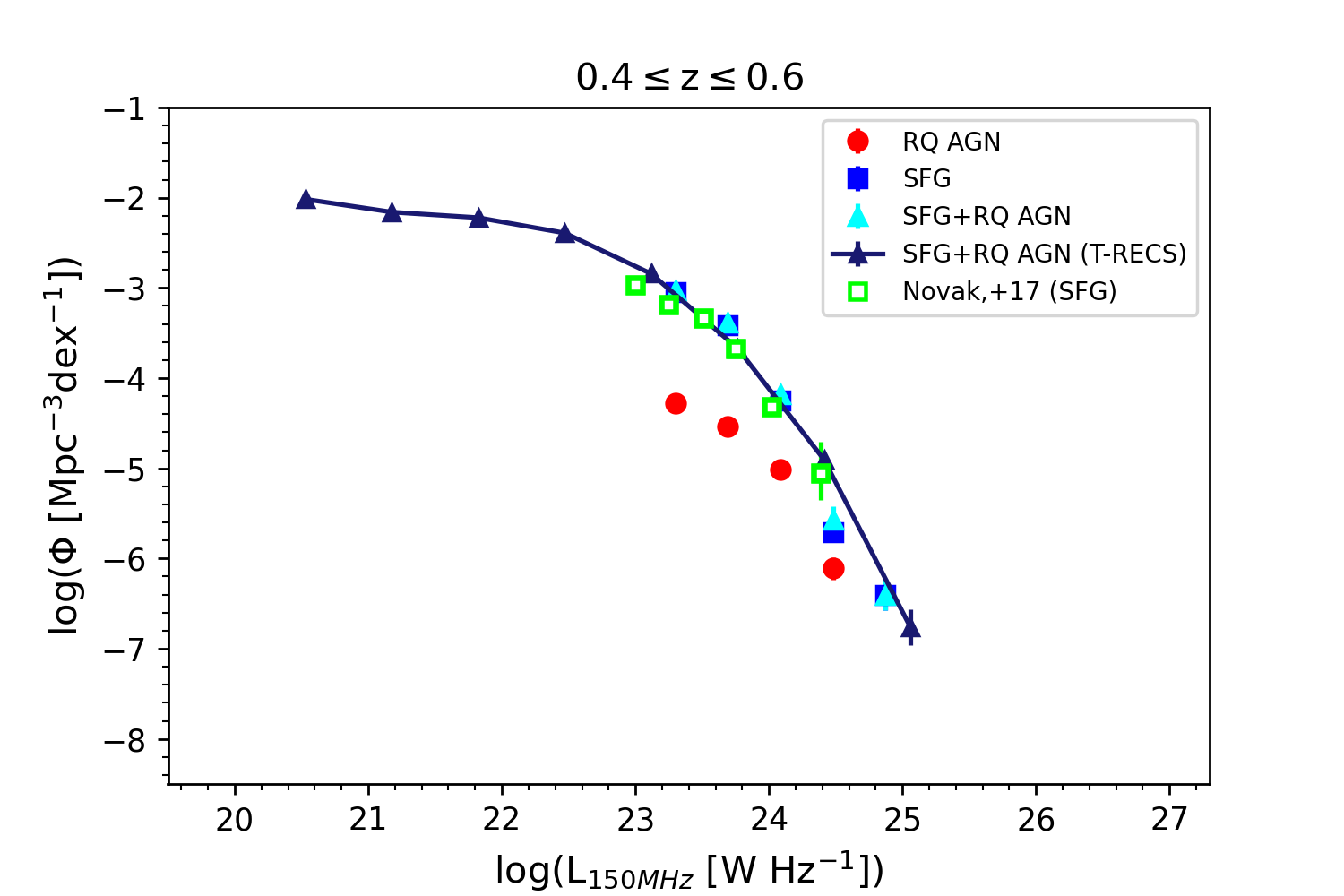}
\includegraphics[width=0.49\textwidth]{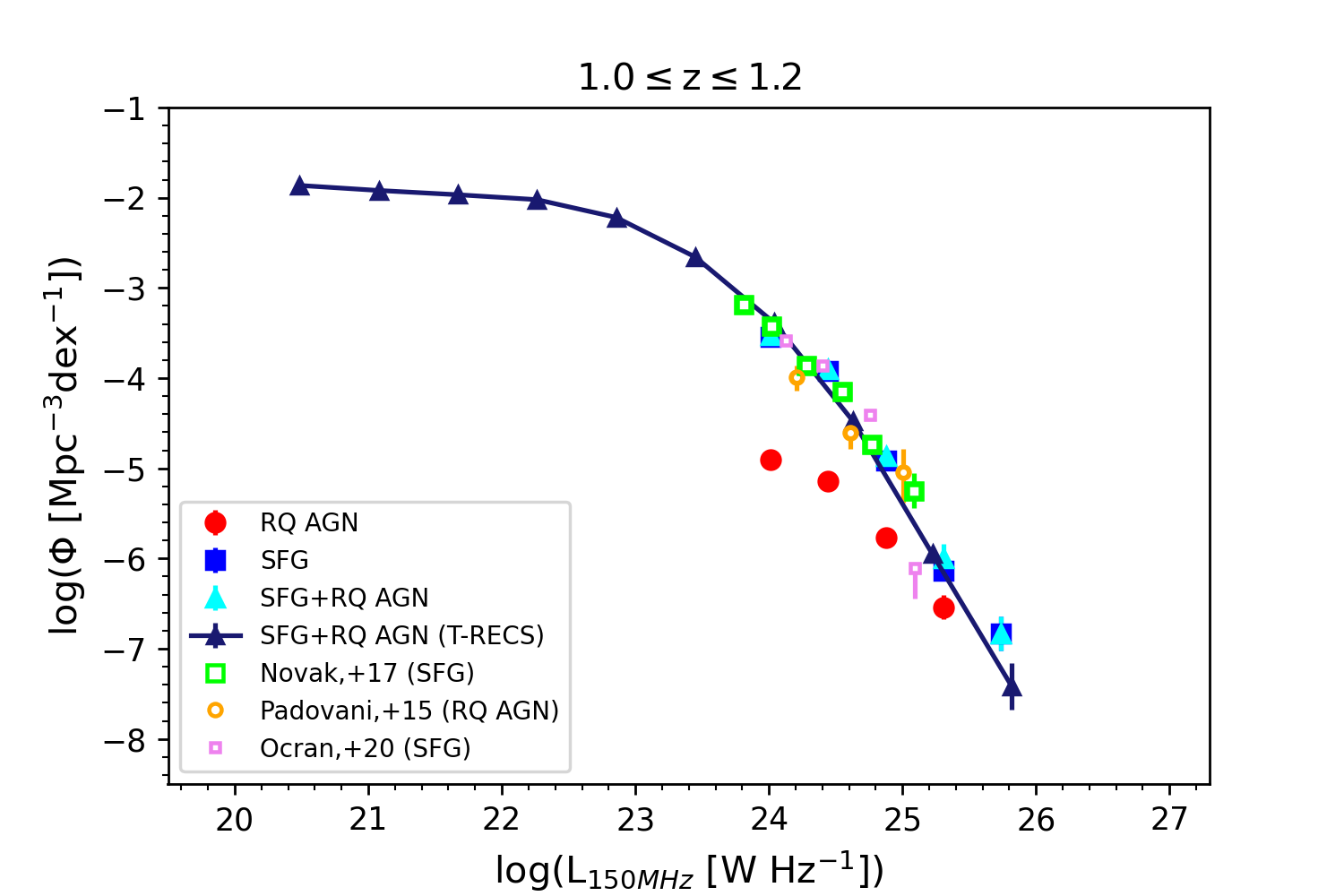}
\includegraphics[width=0.49\textwidth]{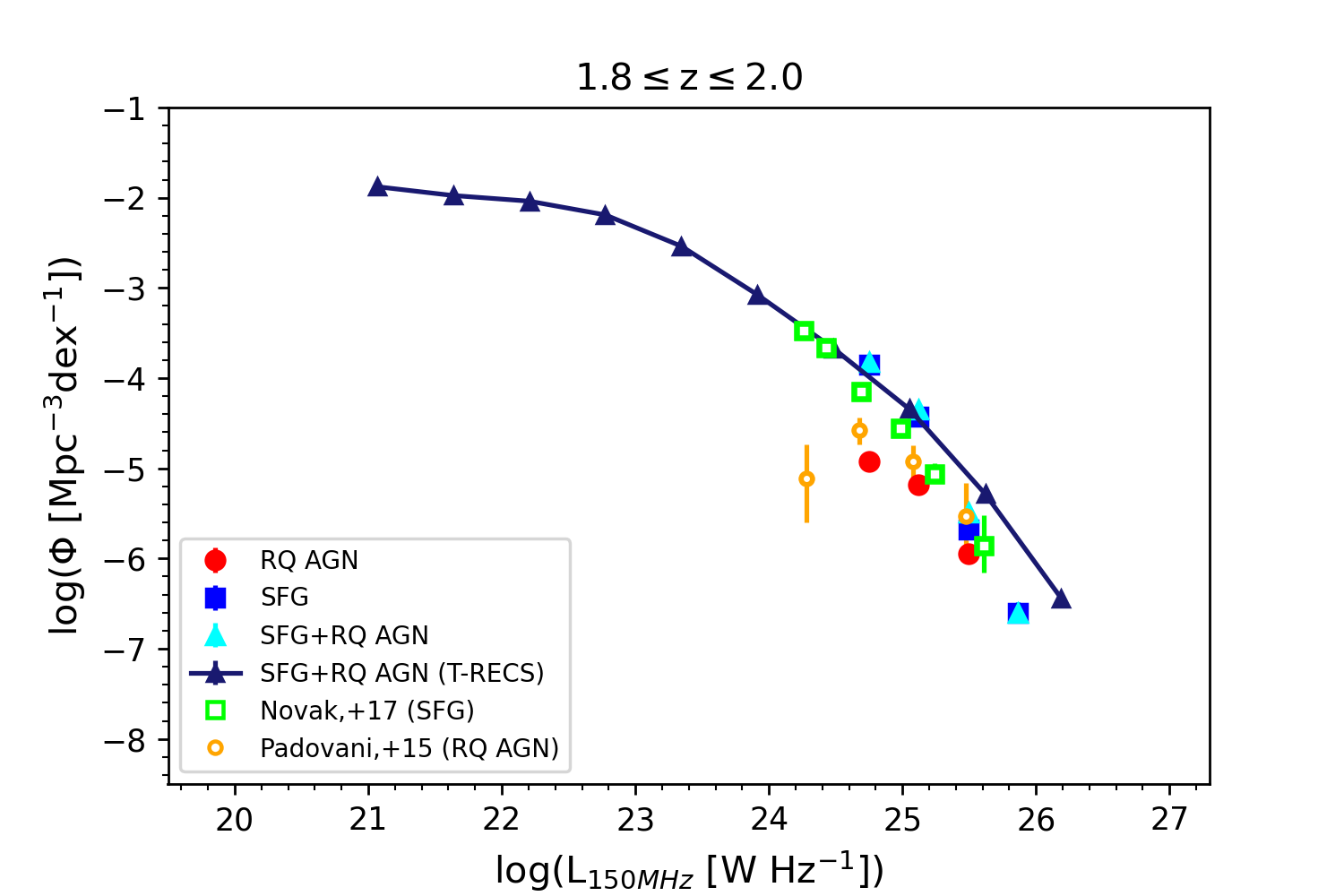}
\includegraphics[width=0.49\textwidth]{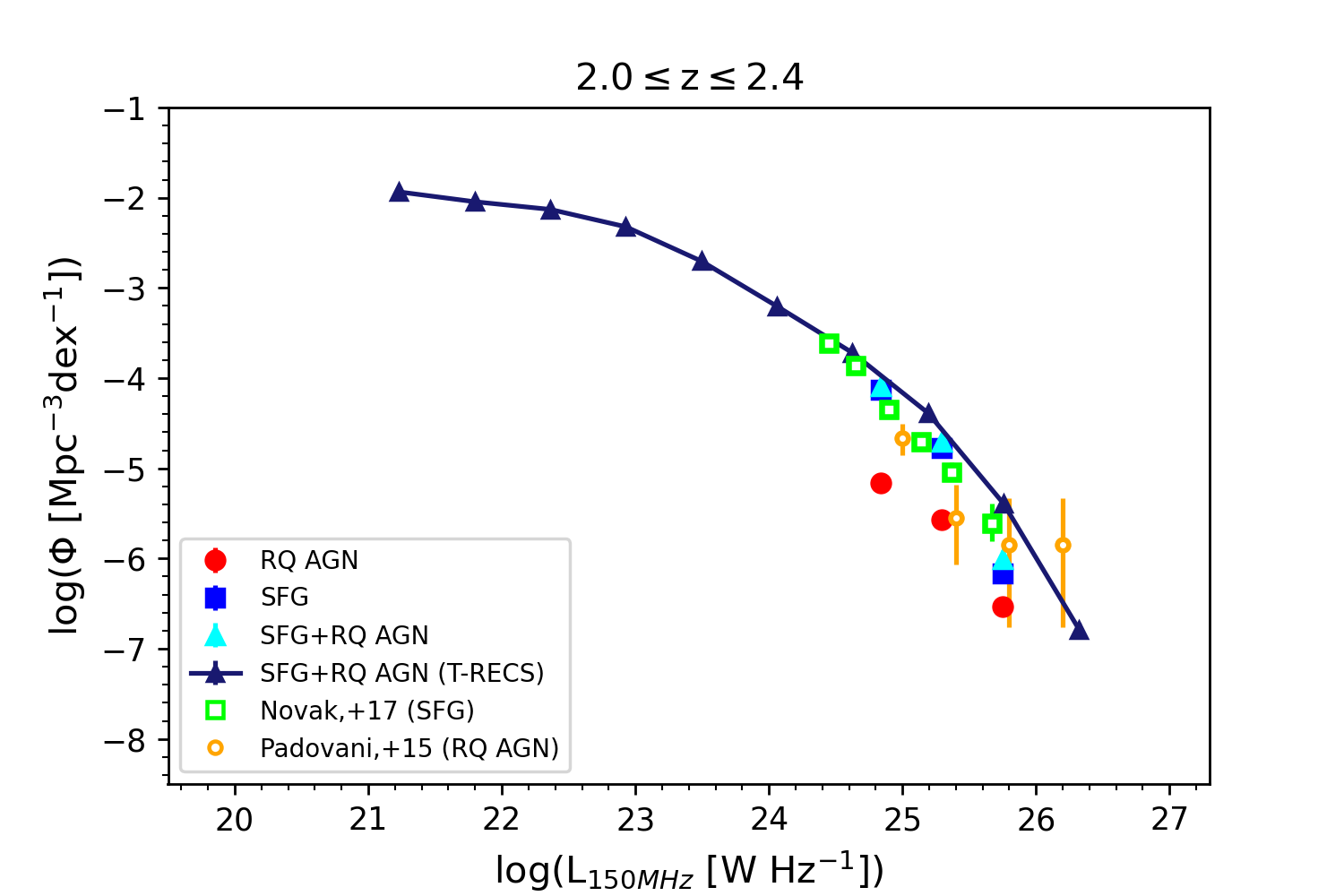}
\includegraphics[width=0.49\textwidth]{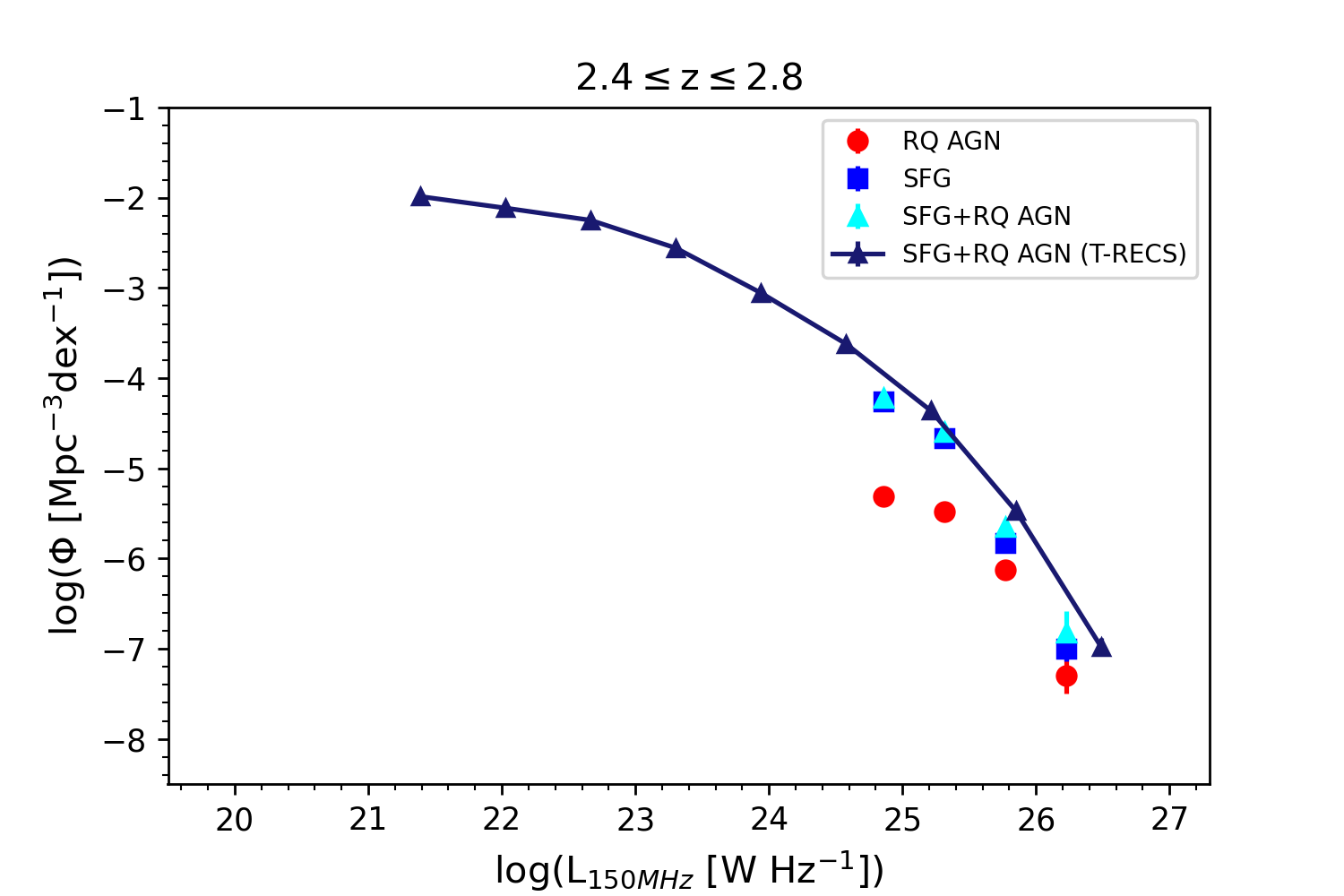}
\caption{Comparison of our estimates of the 150\,MHz luminosity functions (LFs) of non-radio excess sources (SFGs and RQ AGN)
at representative redshifts with those of SFGs (including RQ AGN) from the T-RECS
simulations (\citealt{Bonaldi2019}). Our  LFs are also compared with
the LF estimates at 1.4\,GHz by \citet{Butler2019}, \citet{Novak17}, \citet{Padovani2015} and \citet{MauchSadler2007}, and at 610\,MHz by \citet{Ocran2020}, all scaled to 150\,MHz using a
spectral index $\alpha=-0.73$. The local ($0.05\leq z \leq 0.2$) LFs are also compared with the \citet{Wilman2008} model for SFGs and RQ AGN. The error bars are the quadratic sums of Poisson uncertainties and of sample variance; they are generally smaller than the symbols. The real uncertainties are much larger, especially at $z>1$ where the quality of photometric redshifts increasingly worsens,  but are difficult to quantify. }
 \label{fig:RLF_SFG}
  \end{center}
\end{figure*}

\subsection{Radio luminosity functions}\label{subsec:RLF}

The luminosity of a source with flux density $S_\nu$ is
\begin{equation}\label{eq:Lnu}
L_\nu=\frac{4\pi d_L^2 S_\nu}{K(L_\nu,z)},
\end{equation}
$d_L$ being the luminosity distance and $K(L_\nu,z)$ the K-correction
\begin{equation}
K(L_\nu,z)=\frac{(1+z)L[\nu(1+z)]}{L(\nu)}.
\end{equation}
We have assumed simple power-law spectra, $S_\nu = \nu^{\alpha}$, so that
$K = (1+z)^{1+\alpha}$. We have used the median value of the 1.4\,GHz--150\,MHz
spectral index (-0.73) derived from non-RL AGN sources (SFGs and RQ AGN) with counterparts in the WSRT survey (see Sect.\,\ref{subsec:spectral_indices})

The luminosity function (LF) in the $k$-th redshift bin was derived using the
$1/V_{\rm max}$ method (\citealt{Schmidt1968}):
\begin{equation}
\frac{dN(L_j,z_k)}{d\log L}=\frac{1}{\Delta \log L}\sum_{i=1}^{N_{j}} \frac{w_{i}}{V_{{\rm max},i}(z_k)}
\end{equation}
where $z_k$ is the bin center and the sum is over all the $N_{j}$ sources with
luminosity in the range [$\log L_{j}-\Delta\log L/2$, $\log L_{j}+\Delta\log
L/2$] within the redshift bin. We chose $\Delta z=0.2$ for $z < 2$ and $\Delta
z=0.4$ for $z > 2$. These redshift bins are large enough to have sufficient
statistics and narrow enough to allow us to neglect evolutionary corrections
within the bin. The values of $\Delta\log L$ are different in different
redshift bins (see Table~\ref{tab:RLF}); they are larger in the poorly
populated luminosity ranges.

The $V_{{\rm max},i}$ is the comoving volume within the solid angle of the
survey enclosed between the lower limit, $z_{\rm min}$, of the bin and the
minimum between the upper limit, $z_{\rm max}$, and the maximum redshift at
which the source is above the radio flux-density limit that we assumed, i.e. $S_{\rm
lim}=145\,\mu$Jy. 

The weights $w_i$ include corrections for incompleteness in the radio source catalogue due to the variable rms noise across the survey area, in classification and in redshift. Incompleteness in source counts was corrected
using the factors estimated by \citet{Mandal2020}, as explained in Sect.\,\ref{sect:data}. \citet{Mandal2020} estimated the resolution bias correction based on an
integrated source size distribution, which is the result of the relative contribution
of (more compact) SFGs and (more extended) RL AGN as a function of flux
density. We caution that in principle this may result in not fully appropriate
corrections to SFG LFs at flux densities where RL AGN dominate.  We notice
however that significant resolution bias corrections only apply
at flux densities fainter than 0.3-0.4 mJy, where the  radio catalogue  is
strongly dominated by SFGs and hence the integrated source size distribution is
SFG-like. At higher flux densities the corrections are negligible.

The completeness in the classification or in redshift for each bin of $\log(S_{150\,\rm MHz})$  was estimated as the ratio between the number of classified sources, or of sources with reliable redshift, and the total number of sources in such a bin. While acknowledging that this procedure is not ideal, we note that our results are very weakly sensitive to corrections for these incompletenesses since $\approx 97.6\%$ of our sources are optically identified and $\approx 96.9\%$ have either spectroscopic or reliable photometric redshift (see Sect.\,\ref{sect:data}).

The Poisson error on $dN(L_j,z_k)/d\log L$ was estimated as:
\begin{equation}
\sigma_{j,k} = \left\{\sum_{i=1}^{N_{j}} \frac{w_{i}^{2}}{[V_{{\rm max},i}(z_k)]^{2}}\right\}^{1/2}.
\end{equation}
Our estimates of the 150\,MHz LFs for SFGs and RQ AGN are listed in
Table\,\ref{tab:RLF}. Figure\,\ref{fig:RLF_SFG} displays such LFs for some
representative redshifts and compares them with several estimates found in the literature.

The error bars take into account also the sample variance, i.e. the field-to-field variations of the source number density. According to eq.~(8) of \citet{DeZotti2010} $(\delta N/N)_v^2=2.36\times 10^{-3}  (\Omega/\hbox{deg}^2)^{-0.4}$, $N$ being the mean number density of sources in the bin and $\Omega$ the survey area ($\Omega = 10.28\,\hbox{deg}^2$ in our case). The global error is thus the sum in quadrature of Poisson and sample variance errors.

Our sample has allowed us to determine the luminosity functions at 150\,MHz of
both SFGs and RQ AGN  in the local universe ($z\simeq 0.1$) over about four decades
in luminosity. The results are generally in good agreement with the local luminosity
function at 1.4\,GHz measured by \citet{MauchSadler2007} and, specifically for the SFG population, with the  \citet{Butler2019}, \citet{Novak17} and \citet{Ocran2020} observational estimates. Also good is the agreement with the T-RECS \citep{Bonaldi2019} and the \citet{Wilman2008} simulations of SFGs. However, below the knee luminosity our local LF is somewhat steeper than the \citet{MauchSadler2007} and T-RECS ones. This matches the slightly higher sub-mJy  counts (the bump between 0.25 and 0.8\,mJy) measured by \citet{Mandal2020} compared to the T-RECS simulations. The deficiency of simulated sources is mostly associated to low redshifts.

Our estimates of the RQ AGN LFs are generally lower than those by \citet{Padovani2015} and than the simulations by \citet{Wilman2008}.
We agree with \citet{Padovani2015} and with \citet{Bonato2020} that the RQ AGN fraction is higher at the highest radio luminosities ($\log(L_{150\,\rm MHZ}/\hbox{W}\,\hbox{Hz}^{-1})\simgt 24$). This might mean that RQ AGN are associated to the most powerful starbursts, although some residual AGN contamination cannot be ruled out. Our estimates reach down to the knee luminosity up to $z\simeq 0.5$. At higher redshifts our sample probes only the
highest radio luminosities. The consistency with the T-RECS simulations remains
good although there are hints of incompleteness in our lowest luminosity bins.

\begin{figure*}
\begin{center}
\includegraphics[width=0.49\textwidth]{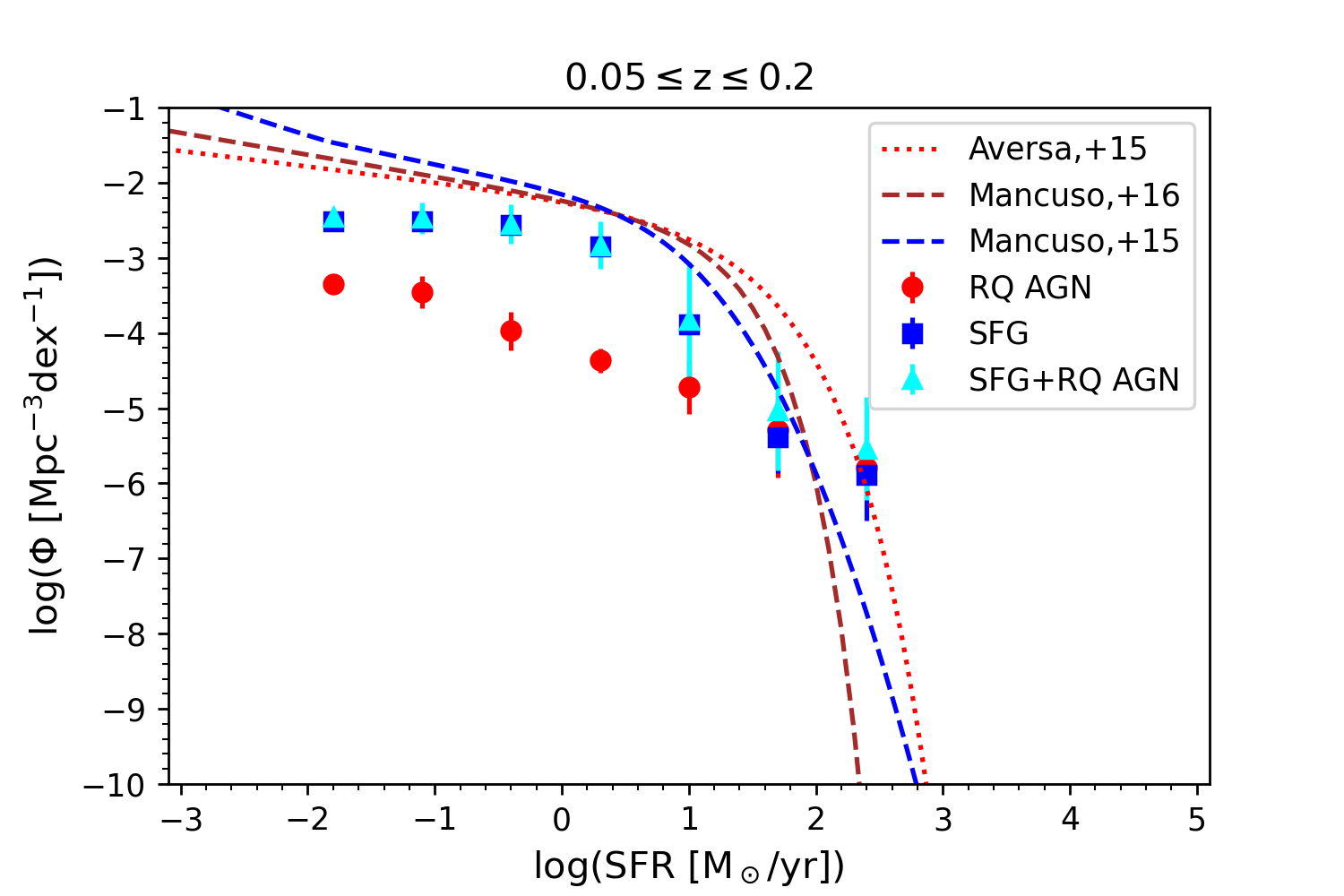}
\includegraphics[width=0.49\textwidth]{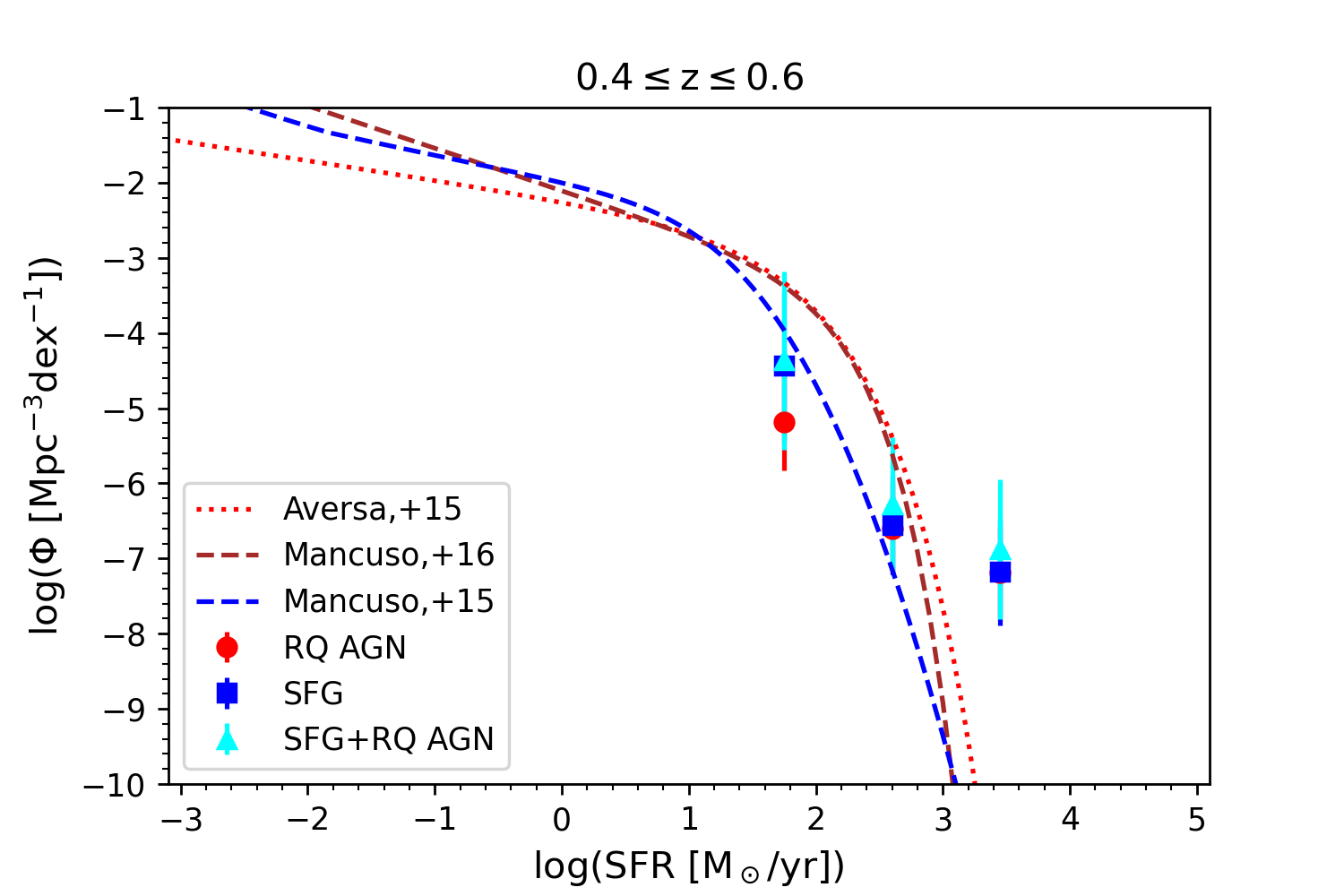}
\includegraphics[width=0.49\textwidth]{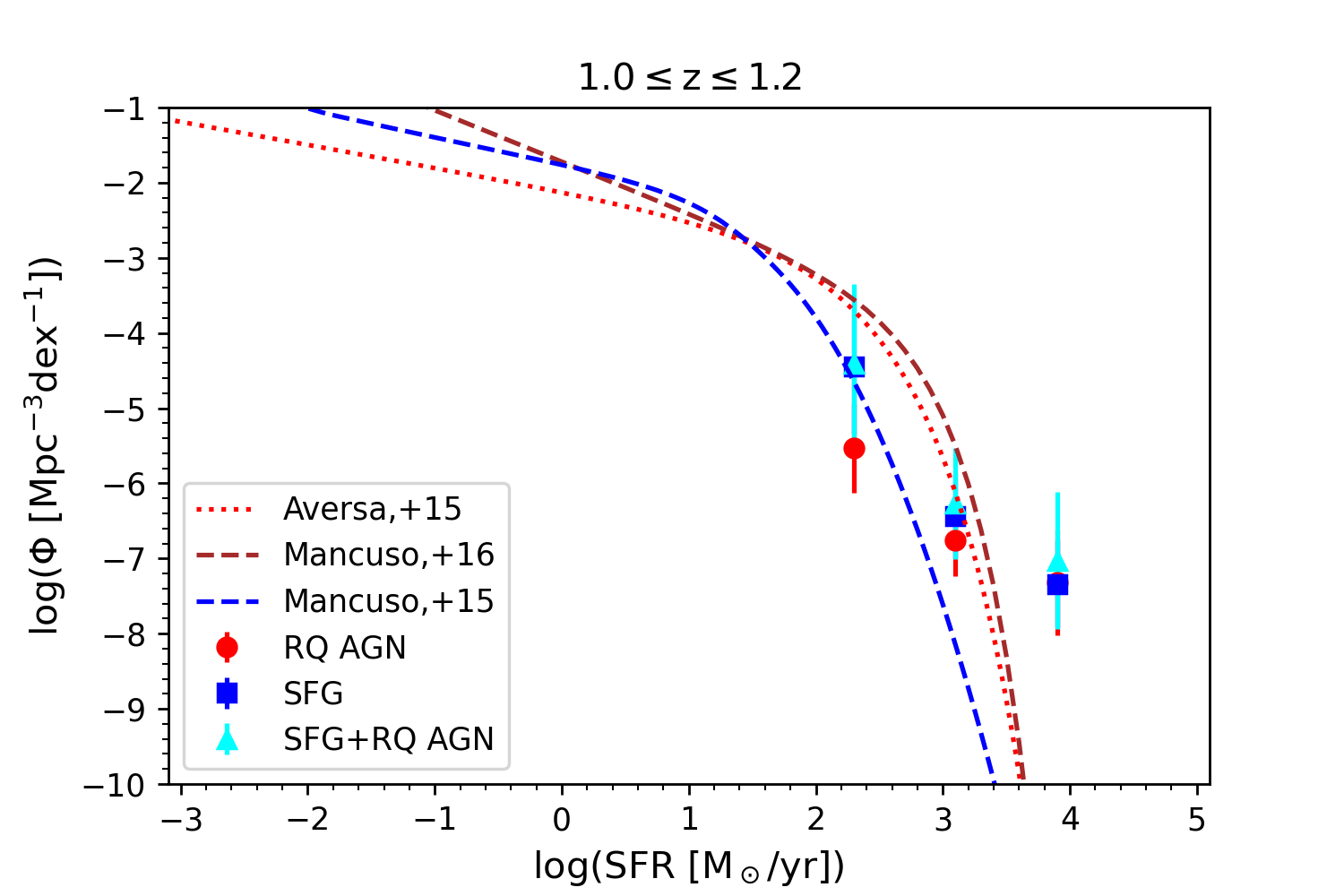}
\includegraphics[width=0.49\textwidth]{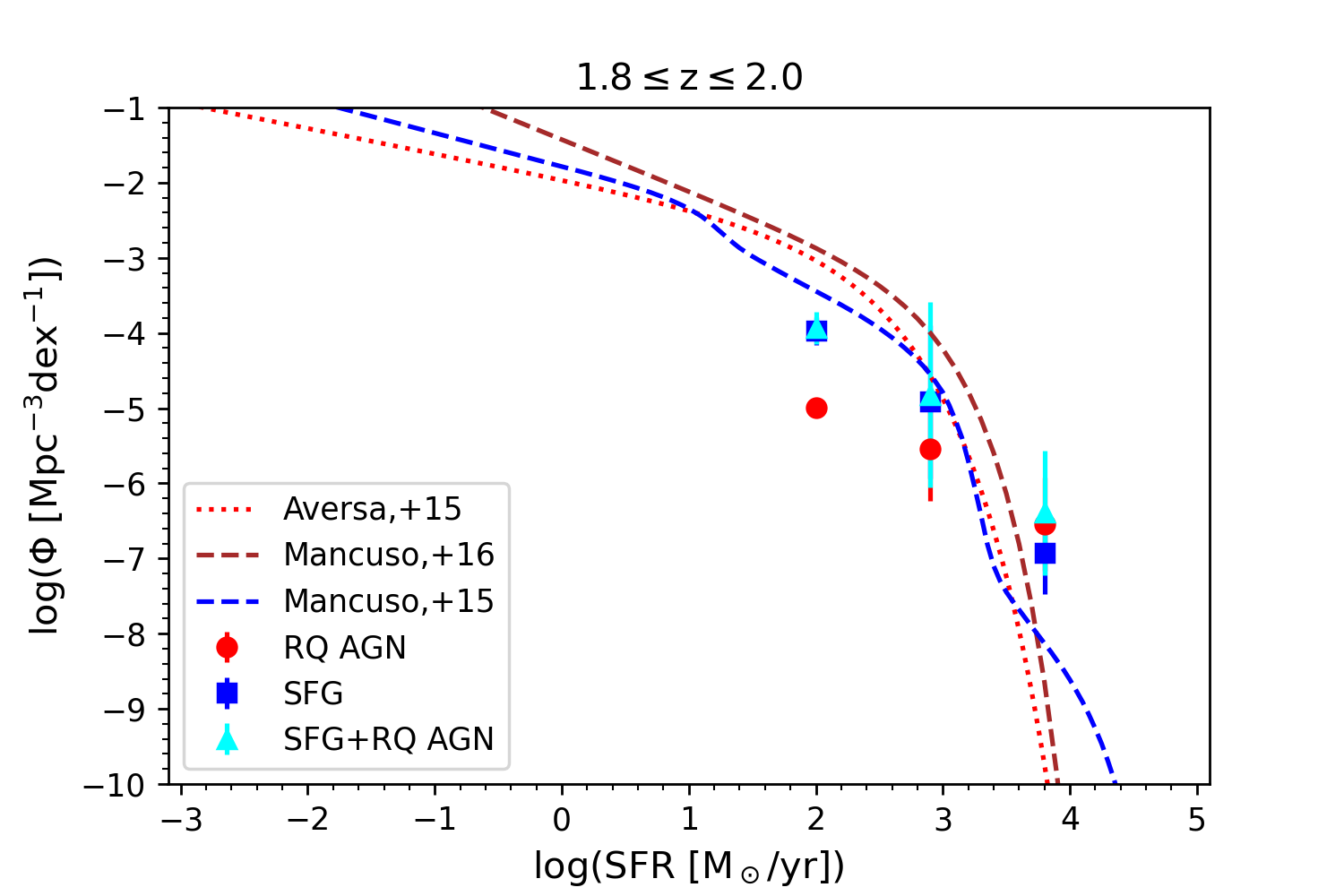}
\includegraphics[width=0.49\textwidth]{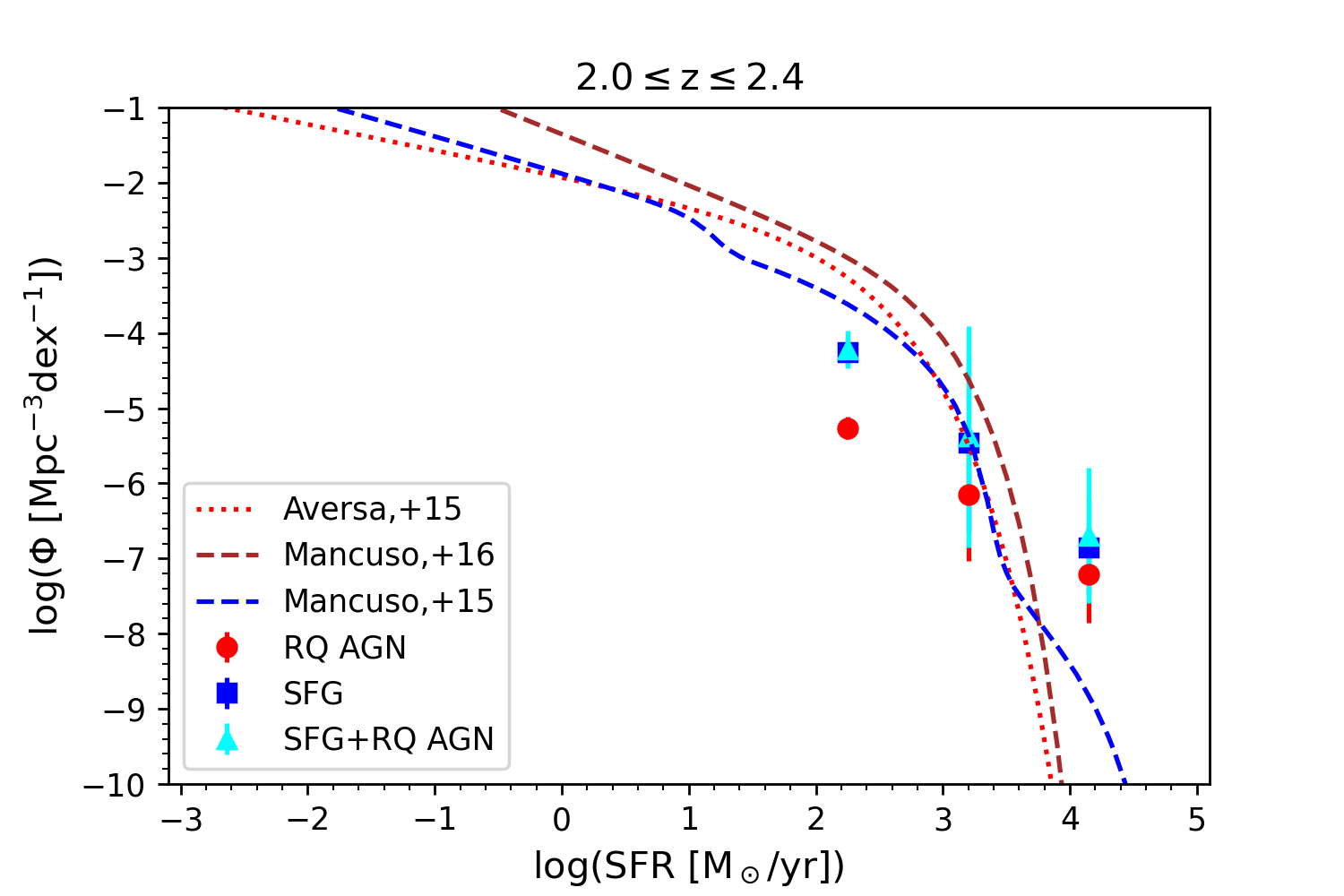}
\includegraphics[width=0.49\textwidth]{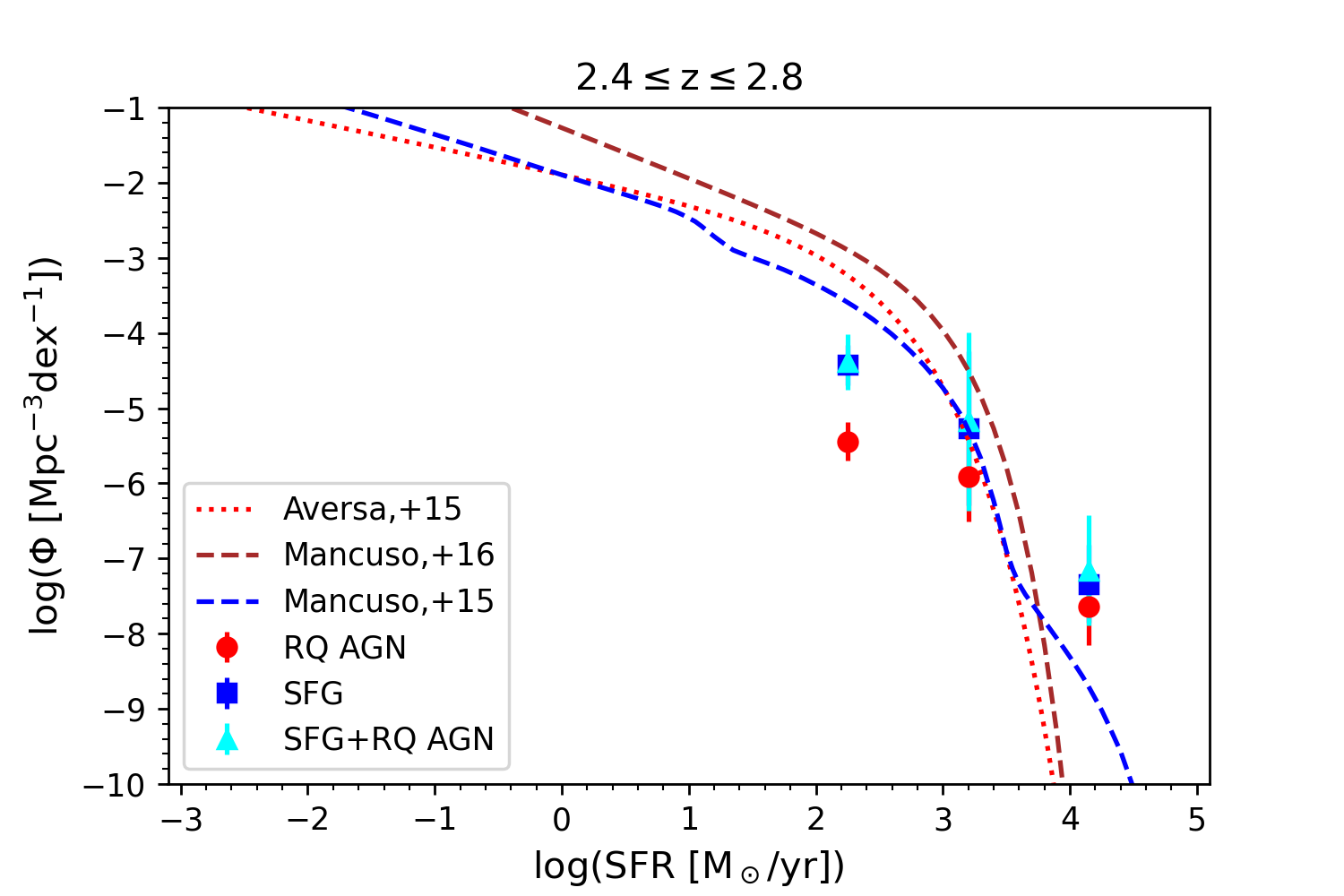}
\caption{Comparison of our observational estimates of the SFR functions with the
theoretical ones by \citet[][blue dashed lines]{Mancuso2015b}, \citet[][red dotted lines]{Aversa2015}
and \citet[][brown dashed lines]{Mancuso2016}. The lines can be seen as representations of observational estimates of the SFR functions based on combinations of classical UV and FIR star-formation tracers (see text). They indicate that the LoTSS data allow us to measure space densities of typical high-$z$ star-forming galaxies,  with $\hbox{SFR}\sim 100\,M_\odot\,\hbox{yr}^{-1}$,  up to $z\simeq 2.5$. The two inflection points, at  $\log(\hbox{SFR}/M_\odot\,\hbox{yr}^{-1})\simeq 1$ and $\simeq 3.5$, of the \citet{Mancuso2015b} line for $z>1.5$ correspond to the transition from the dominance of late-type galaxies to that of star-forming protospheroids and to the dominance of strongly lensed galaxies, respectively. The error bars are the quadratic sums of Poisson uncertainties and of sample variance; they are generally smaller than the symbols. The real uncertainties are much larger, especially at $z>1$ where the quality of photometric redshifts increasingly worsens, but difficult to quantify. }
 \label{fig:SFRf}
  \end{center}
\end{figure*}

\subsection{Star-formation rate functions}\label{subsec:SFRf}

The SFR functions of SFGs and RQ AGN were computed by convolving their radio
LFs with the distributions of the $\hbox{SFR}/L_{150\,\rm MHz}$ ratios at  given
$L_{150\,\rm MHz}$. We used a Monte Carlo approach,
carrying out 100,000 simulations to sample the distribution of the $\hbox{SFR}/L_{150\,\rm MHz}$ ratios as a function of $L_{150\,\rm MHz}$. The distribution was modeled as a Gaussian with a mean given by the $\log(L_{150\,\rm MHz})$--$\log(\hbox{SFR})$ relation by \citet{Best2020}; the dispersion around the mean relation was found to be 0.282.

The results for each population and their sum are listed in
Table\,\ref{tab:SFRf} and displayed in Fig.~\ref{fig:SFRf} for a set of
representative redshifts. The effective numbers of sources given in the table are the number of sources corresponding to the central values of the SFR function: since the latter was determined via Monte Carlo simulations, there is no unique number of sources in each bin. Estimates of the SFR functions of SFGs in their two highest redshift bins ($\langle z \rangle = 3.7$ and 4.8) were worked out by \citet{Novak17}.

Figure~\ref{fig:SFRf} also shows, for comparison, analytic fits of estimates of
the global SFR functions by \citet{Mancuso2015b},  \citet{Aversa2015} and
\citet{Mancuso2016}, all based on FIR/sub-mm, optical and UV data. In fact, the lines in this figure can be seen as representations of data from such spectral regions.

\citet{Mancuso2015b} complemented the SFR functions yielded by the
\citet{Cai2013} model that fitted a broad variety of IR data on dust-obscured
star formation over a wide redshift range, with a parametric model for the
unobscured star formation, successfully tested against the observed H$\alpha$
and UV luminosity functions.

\citet{Aversa2015} described the bolometric luminosity due to star formation as
the sum of UV plus IR luminosities in such proportions that the observed LFs in
both bands at all redshifts are reproduced. Whenever insufficient data were
available, the SFR functions were constrained using the continuity equation to
link them to the halo mass function.

\citet{Mancuso2016} performed educated minimum--$\chi^2$ fits of IR and
dust-corrected data assuming a Schechter shape for the SFR functions, with
redshift-dependent parameters.

Our observational estimates are in reasonably good agreement with the results by
\citet{Mancuso2015b} but are generally below those by \citet{Mancuso2016} and
even more below those by \citet{Aversa2015}. The inflection points of the
\citet{Mancuso2015b} function at $z> 1.5$ correspond to the changes of the
dominant population according to the \citet{Cai2013} model. At low $z$ the
dominant population is made of normal and starburst late-type galaxies, while
at $z> 1.5$ proto-spheroidal galaxies and bulges in the process of forming the
bulk of their stars take over above $\log(\hbox{SFR}/M_\odot\,\hbox{yr}^{-1})\simeq 1$ (first inflection point). Strongly lensed galaxies account for the highest (apparent) SFRs, above $\log(\hbox{SFR}/M_\odot\,\hbox{yr}^{-1})\simeq 3.5$ (second inflection point). Figure~\ref{fig:SFRf} shows that our sources do not reach the extreme values of the SFR where strongly lensed galaxies are predicted to dominate. Their maximum SFRs are in the range of unlensed hyperluminous infrared galaxies (HyLIRGs) with $\hbox{SFR}\simgt 1000\,M_\odot\,\hbox{yr}^{-1}$ \citep{Fu2013, Ivison2013}. However source blending \citep[e.g.][]{Hayward2013} and AGN contributions to the IR luminosity used to derive the SFR (a large fraction of the highest SFR galaxies host a RQ AGN, see Fig.~\ref{fig:SFRf}) can also have a role.

It is remarkable that radio data alone have allowed us to derive SFR functions in broad agreement with those based on a complex combination of UV to optical to IR data. The differences can be accounted for by the still substantial uncertainties on the calibration of recipes for deriving the SFR from data in the various wavebands. The comparison with representations of data from classical SFR measures shows that the LoTSS survey is deep enough to allow the determination of space densities
of typical star-forming galaxies (with SFR at the knee of the function,
$\hbox{SFR}\sim 100\,M_\odot\,\hbox{yr}^{-1}$) up to $z\simeq 2.5$. SFGs not
hosting AGN dominate the global SFR functions although the contribution of AGN
hosts increases with increasing SFR. 

\section{Conclusions}\label{sec:conclusions}

We have exploited the unprecedented sensitivity of LoTSS Deep Fields survey of the Lockman Hole field to investigate the relation between radio emission and SFR in SFGs. The LH field is
exceptionally well suited for this purpose given the wealth of multi-frequency
data available on radio-detected sources. Of the 31,162 sources in the final LH
cross-matched catalogue, 97.6\% have multi-wavelength counterparts and 96.9\% have either
spectroscopic or high-quality photometric redshift.

We have adopted source classification and estimates of SFRs and stellar masses made by \citet{Best2020} using four different SED fitting methods, optimized  for either purely star-forming galaxies or for galaxies hosting an AGN. This has
put the results on more solid grounds compared to previous analyses generally
relying on just one method.

A comparison with the analysis by \citet{Bonato2020} of deep WSRT observations
of a fraction of the field has shown that a significant fraction ($\simeq
26\%$) of sources classified as RL AGN by \citet{Bonato2020} have a flatter than average 1.4\,GHz--150\,MHz spectral index (median $-0.55$, to be compared with a median $-0.72$ for confirmed RL AGN). As a consequence, not all of them show radio excess at 150\,MHz and were classified as SFGs or RQ AGN by \citet{Best2020}. This indicates that the conclusion related to the origin of the dominant radio emission (nuclear activity or star formation) may be different at different frequencies, in addition to the effect of different photometry used for classification from SED fitting.


We found a slight trend, weaker than that reported by \citet{Smith2020} and \citet{Delvecchio2020},
towards a lower $\log(\hbox{SFR}/L_{150\,\rm MHz})$ ratio for higher
$M_\star$. We interpret the differences in terms of the effect of the different primary selections and note that we get good agreement with \citet{Smith2020} for objects with typical SFR and $M_\star$, the least affected by selection effects. We
confirm the decrease of this ratio with increasing redshift reported in previous papers \citep{Magnelli2015, Delhaize2017, CalistroRivera2017}, except for the most massive SFGs. Our results are consistent with the evolution of
the $\log(L_{150\,\rm MHz}/\hbox{SFR})$ being mostly driven by its increase
with stellar mass as argued by \citet{Smith2020} and \citet{Delvecchio2020}.

Our data show, for $z>0.5$, a decrease of the sSFR with increasing $M_\star$, similar to that reported by \citet{Bourne2017}, although with a hint of an excess sSFR of the lowest stellar masses. We do not see the flattening of the slope of the $\log(\hbox{sSFR})$ vs. $\log(M_\star)$ relation at low $M_\star$ reported by \citet{Whitaker2014}. This suggests   a higher efficiency of the radio selection at detecting early phases of galaxy evolution, when the stellar masses were still relatively low or effects of selection biases -- see notes made in earlier sections. Also the redshift evolution of the amplitude of the  $\log(\hbox{sSFR})$ vs. $\log(M_\star)$ relation is consistent with the results by \citet{Bourne2017} but is substantially weaker than found by several other studies.

SFGs in our sample do not show a clear correlation between SFR and stellar
mass, i.e. of the so-called ``galaxy main sequence'', a situation similar to
that found for sub-mm selected galaxies. The view of the distribution of galaxies in the $\hbox{SFR}$--$M_\star$ plane from the star-formation perspective, presented by these selections, is complementary to that coming from the optical/near-IR selections which emphasize $M_\star$. The uniformity of the distribution argues against different star-formation modes for ``main sequence'' and ``off-sequence'' galaxies.

We have derived luminosity functions at 150\,MHz of both SFGs and RQ AGN at
various redshifts. Our results for the SFG LFs are in good agreement with the T-RECS and \citet{Wilman2008} simulations, with the local
luminosity function by \citet{MauchSadler2007} and with the  \citet{Butler2019}, \citet{Novak17} and \citet{Ocran2020} estimates. Our estimates reach down to the knee luminosity up to $z\simeq 0.5$. Deeper radio surveys are necessary to determine the space density of galaxies with
typical radio luminosity at higher $z$. Our LFs of RQ AGN are somewhat below those by \citet{Padovani2015} and the \citet{Wilman2008} model.

We have also presented explicit estimates of SFR
functions of SFGs and RQ AGN derived from radio survey data. Our estimates
are in good agreement with the results by \citet{Mancuso2015b} based on
FIR/sub-mm/optical/UV data but are generally below those by
\citet{Mancuso2016} and even more below those by \citet{Aversa2015}. SFR functions are dominated at all redshifts by pure SFGs, but with a large contribution of RQ AGN at the highest SFRs.

In conclusion, the new data are improving our understanding of the radio/SFR relation of SFGs, but increasing complexities, such as dependencies of the relation on additional parameters like the stellar mass and redshift,  are emerging. Better data, i.e. more extensive spectroscopy and deeper multi-wavelength data over wider areas, are necessary before the calibration of the radio/SFR relation and its dependence on other parameters  can
be accurately assessed.

\begin{acknowledgements}
We are grateful to the anonymous referee for a careful reading of the manuscript and many useful comments. MB and IP acknowledge support from INAF under PRIN SKA/CTA ``FORECaST'' and PRIN MAIN STREAM ``SAuROS''. MB acknowledges support from the
Ministero degli Affari Esteri e della Cooperazione Internazionale - Direzione Generale per la Promozione del Sistema Paese Progetto di Grande Rilevanza ZA18GR02 and from the South African Department of Science and Technology's National Research Foundation (DST-NRF Grant Number 113121). PNB acknowledges support from STFC through grants ST/R000972/1 and ST/V000594/1. KM has been supported by the National Science Centre (UMO-2018/30/E/ST9/00082). RK acknowledges support from the Science and Technology Facilities Council (STFC) through an STFC studentship via grant ST/R504737/1.
\end{acknowledgements}

\bibliographystyle{aa} 
\bibliography{biblio.bib}



\begin{table*}
\centering
\caption{Estimates of the 150\,MHz LF, $\Phi=dN/(d\log L\, dV) [\hbox{Mpc}^{-3}]$, for SFGs,
RQ AGN and the sum of the two populations at redshifts of up to $\simeq 4$. Errors are the quadratic sum of Poisson uncertainties and sample variance. On the right of the luminosity functions we give the (uncorrected) number, N, of sources in each bin.}
\begin{tabular}{ccccccc}
\hline
\hline
$\log L$ & $\log(\Phi_{\rm SFG + RQ AGN})$ & N$_{\rm SFG + RQ AGN}$ & $\log(\Phi_{\rm SFG})$ & N$_{\rm SFG}$ & $\log(\Phi_{\rm RQ\, AGN})$ & N$_{\rm RQ\,AGN}$  \\
$[\hbox{W}\,\hbox{Hz}^{-1}]$ & $[\hbox{Mpc}^{-3}\,\hbox{dex}^{-1}]$ & & $[\hbox{Mpc}^{-3}\,\hbox{dex}^{-1}]$ & &
$[\hbox{Mpc}^{-3}\,\hbox{dex}^{-1}]$ & \\
\hline
\hline
\multicolumn{7}{c}{z = 0.05$-$ 0.2 } \\
21.36  &  -2.44 $\pm$ 0.11  &  38  &  -2.51 $\pm$ 0.05  &  33  &  -3.31 $\pm$ 0.11  &  5  \\
21.85  &  -2.51 $\pm$ 0.08  &  214  &  -2.53 $\pm$ 0.02  &  203  &  -3.84 $\pm$ 0.07  &  11  \\
22.34  &  -2.59 $\pm$ 0.06  &  538  &  -2.60 $\pm$ 0.02  &  524  &  -4.19 $\pm$ 0.06  &  14  \\
22.84  &  -2.87 $\pm$ 0.07  &  331  &  -2.89 $\pm$ 0.02  &  321  &  -4.39 $\pm$ 0.07  &  10  \\
23.33  &  -3.67 $\pm$ 0.09  &  52  &  -3.73 $\pm$ 0.03  &  46  &  -4.61 $\pm$ 0.09  &  6  \\
23.82  &  -5.08 $\pm$ 0.15  &  2  &  -5.08 $\pm$ 0.15  &  2  &  $-$ & $-$  \\
\hline
\multicolumn{7}{c}{z = 0.2 $-$ 0.4 } \\
22.70  &  -2.88 $\pm$ 0.04  &  617  &  -2.90 $\pm$ 0.02  &  591  &  -4.28 $\pm$ 0.04  &  26  \\
23.10  &  -2.98 $\pm$ 0.03  &  1051  &  -3.01 $\pm$ 0.01  &  986  &  -4.19 $\pm$ 0.03  &  65  \\
23.51  &  -3.43 $\pm$ 0.04  &  450  &  -3.45 $\pm$ 0.02  &  427  &  -4.72 $\pm$ 0.04  &  23  \\
23.91  &  -4.39 $\pm$ 0.09  &  49  &  -4.43 $\pm$ 0.03  &  45  &  -5.48 $\pm$ 0.09  &  4  \\
24.31  &  -5.30 $\pm$ 0.19  &  6  &  -5.37 $\pm$ 0.08  &  5  &  -6.07 $\pm$ 0.17  &  1  \\
\hline
\multicolumn{7}{c}{z = 0.4 $-$ 0.6 } \\
23.30  &  -3.02 $\pm$ 0.03  &  1417  &  -3.05 $\pm$ 0.01  &  1330  &  -4.27 $\pm$ 0.02  &  87  \\
23.69  &  -3.38 $\pm$ 0.03  &  1027  &  -3.41 $\pm$ 0.01  &  955  &  -4.54 $\pm$ 0.02  &  72  \\
24.09  &  -4.18 $\pm$ 0.04  &  171  &  -4.25 $\pm$ 0.02  &  146  &  -5.02 $\pm$ 0.04  &  25  \\
24.48  &  -5.57 $\pm$ 0.14  &  7  &  -5.71 $\pm$ 0.08  &  5  &  -6.11 $\pm$ 0.12  &  2  \\
24.87  &  -6.40 $\pm$ 0.17  &  1  &  -6.40 $\pm$ 0.17  &  1  &  $-$ & $-$  \\
\hline
\multicolumn{7}{c}{z = 0.6 $-$ 0.8 } \\
23.60  &  -3.20 $\pm$ 0.04  &  1584  &  -3.21 $\pm$ 0.01  &  1533  &  -4.72 $\pm$ 0.03  &  51  \\
24.06  &  -3.69 $\pm$ 0.03  &  935  &  -3.72 $\pm$ 0.01  &  869  &  -4.84 $\pm$ 0.03  &  66  \\
24.53  &  -4.81 $\pm$ 0.06  &  74  &  -4.91 $\pm$ 0.03  &  59  &  -5.50 $\pm$ 0.05  &  15  \\
24.99  &  -5.82 $\pm$ 0.17  &  7  &  -5.97 $\pm$ 0.09  &  5  &  -6.36 $\pm$ 0.14  &  2  \\
\hline
\multicolumn{7}{c}{z = 0.8 $-$ 1.0 } \\
24.00  &  -3.45 $\pm$ 0.03  &  1767  &  -3.47 $\pm$ 0.01  &  1681  &  -4.80 $\pm$ 0.03  &  86  \\
24.55  &  -4.38 $\pm$ 0.05  &  312  &  -4.43 $\pm$ 0.02  &  275  &  -5.30 $\pm$ 0.04  &  37  \\
25.10  &  -6.02 $\pm$ 0.09  &  7  &  -6.02 $\pm$ 0.09  &  7  &  $-$ & $-$  \\
25.66  &  -6.86 $\pm$ 0.24  &  1  &  $-$ & $-$  &  -6.86 $\pm$ 0.24  &  1  \\
\hline
\multicolumn{7}{c}{z = 1.0 $-$ 1.2 } \\
24.01  &  -3.53 $\pm$ 0.04  &  1038  &  -3.54 $\pm$ 0.02  &  994  &  -4.90 $\pm$ 0.03  &  44  \\
24.44  &  -3.90 $\pm$ 0.03  &  834  &  -3.92 $\pm$ 0.01  &  786  &  -5.14 $\pm$ 0.03  &  48  \\
24.88  &  -4.86 $\pm$ 0.06  &  96  &  -4.92 $\pm$ 0.02  &  84  &  -5.77 $\pm$ 0.06  &  12  \\
25.31  &  -6.00 $\pm$ 0.16  &  7  &  -6.14 $\pm$ 0.08  &  5  &  -6.54 $\pm$ 0.13  &  2  \\
25.74  &  -6.83 $\pm$ 0.19  &  1  &  -6.83 $\pm$ 0.19  &  1  &  $-$ & $-$  \\
\hline
\multicolumn{7}{c}{z = 1.2 $-$ 1.4 } \\
24.40  &  -3.96 $\pm$ 0.03  &  615  &  -4.00 $\pm$ 0.02  &  555  &  -5.00 $\pm$ 0.03  &  60  \\
24.80  &  -4.61 $\pm$ 0.04  &  183  &  -4.66 $\pm$ 0.02  &  162  &  -5.55 $\pm$ 0.04  &  21  \\
25.21  &  -5.53 $\pm$ 0.10  &  22  &  -5.61 $\pm$ 0.04  &  18  &  -6.27 $\pm$ 0.09  &  4  \\
25.62  &  -6.86 $\pm$ 0.18  &  1  &  $-$ & $-$  &  -6.86 $\pm$ 0.18  &  1  \\
\hline
\multicolumn{7}{c}{z = 1.4 $-$ 1.6 } \\
24.51  &  -4.09 $\pm$ 0.03  &  459  &  -4.14 $\pm$ 0.02  &  398  &  -4.98 $\pm$ 0.03  &  61  \\
24.90  &  -4.71 $\pm$ 0.04  &  151  &  -4.80 $\pm$ 0.02  &  124  &  -5.46 $\pm$ 0.04  &  27  \\
25.30  &  -5.57 $\pm$ 0.08  &  21  &  -5.72 $\pm$ 0.05  &  15  &  -6.11 $\pm$ 0.07  &  6  \\
25.69  &  -6.58 $\pm$ 0.24  &  2  &  -6.88 $\pm$ 0.17  &  1  &  -6.89 $\pm$ 0.17  &  1  \\
\hline
\multicolumn{7}{c}{z = 1.6 $-$ 1.8 } \\
24.66  &  -3.82 $\pm$ 0.03  &  1069  &  -3.86 $\pm$ 0.01  &  964  &  -4.85 $\pm$ 0.02  &  105  \\
25.13  &  -4.62 $\pm$ 0.04  &  237  &  -4.68 $\pm$ 0.02  &  204  &  -5.47 $\pm$ 0.04  &  33  \\
25.60  &  -5.71 $\pm$ 0.21  &  19  &  -5.73 $\pm$ 0.05  &  18  &  -6.99 $\pm$ 0.20  &  1  \\
\hline
\multicolumn{7}{c}{z = 1.8 $-$ 2.0 } \\
24.75  &  -3.82 $\pm$ 0.03  &  892  &  -3.85 $\pm$ 0.01  &  819  &  -4.93 $\pm$ 0.02  &  73  \\
25.12  &  -4.35 $\pm$ 0.03  &  355  &  -4.42 $\pm$ 0.02  &  302  &  -5.18 $\pm$ 0.03  &  53  \\
25.49  &  -5.49 $\pm$ 0.07  &  26  &  -5.68 $\pm$ 0.04  &  17  &  -5.95 $\pm$ 0.06  &  9  \\
25.86  &  -6.60 $\pm$ 0.11  &  2  &  -6.60 $\pm$ 0.11  &  2  &  $-$ & $-$  \\
\hline
\hline
\end{tabular}
\label{tab:RLF}
\end{table*}

\renewcommand{\thetable}{\arabic{table}}
\addtocounter{table}{-1}
\begin{table*}
\centering
\caption{...continued.}
\begin{tabular}{ccccccc}
\hline
\hline
$\log L$ & $\log(\Phi_{\rm SFG + RQ AGN})$ & N$_{\rm SFG + RQ AGN}$ & $\log(\Phi_{\rm SFG})$ & N$_{\rm SFG}$ & $\log(\Phi_{\rm RQ\, AGN})$ & N$_{\rm RQ\,AGN}$  \\
$[\hbox{W}\,\hbox{Hz}^{-1}]$ & $[\hbox{Mpc}^{-3}\,\hbox{dex}^{-1}]$ & & $[\hbox{Mpc}^{-3}\,\hbox{dex}^{-1}]$ & &
$[\hbox{Mpc}^{-3}\,\hbox{dex}^{-1}]$ & \\
\hline
\hline
\multicolumn{7}{c}{z = 2.0 $-$ 2.4 } \\
24.83  &  -4.09 $\pm$ 0.03  &  1090  &  -4.13 $\pm$ 0.01  &  991  &  -5.17 $\pm$ 0.02  &  99  \\
25.29  &  -4.71 $\pm$ 0.03  &  394  &  -4.77 $\pm$ 0.02  &  340  &  -5.57 $\pm$ 0.03  &  54  \\
25.75  &  -6.01 $\pm$ 0.10  &  20  &  -6.16 $\pm$ 0.05  &  14  &  -6.53 $\pm$ 0.08  &  6  \\
\hline
\multicolumn{7}{c}{z = 2.4 $-$ 2.8 } \\
24.86  &  -4.22 $\pm$ 0.04  &  618  &  -4.26 $\pm$ 0.02  &  563  &  -5.32 $\pm$ 0.03  &  55  \\
25.32  &  -4.60 $\pm$ 0.03  &  486  &  -4.66 $\pm$ 0.02  &  421  &  -5.48 $\pm$ 0.03  &  65  \\
25.77  &  -5.65 $\pm$ 0.07  &  45  &  -5.83 $\pm$ 0.04  &  30  &  -6.13 $\pm$ 0.05  &  15  \\
26.23  &  -6.82 $\pm$ 0.24  &  3  &  -7.00 $\pm$ 0.14  &  2  &  -7.30 $\pm$ 0.20  &  1  \\
\hline
\multicolumn{7}{c}{z = 2.8 $-$ 3.2 } \\
25.01  &  -4.43 $\pm$ 0.04  &  511  &  -4.51 $\pm$ 0.02  &  432  &  -5.23 $\pm$ 0.03  &  79  \\
25.56  &  -4.93 $\pm$ 0.04  &  277  &  -5.06 $\pm$ 0.02  &  203  &  -5.50 $\pm$ 0.03  &  74  \\
26.11  &  -6.20 $\pm$ 0.14  &  15  &  -6.34 $\pm$ 0.07  &  11  &  -6.78 $\pm$ 0.12  &  4  \\
\hline
\multicolumn{7}{c}{z = 3.2 $-$ 3.6 } \\
25.52  &  -4.85 $\pm$ 0.04  &  332  &  -5.06 $\pm$ 0.02  &  205  &  -5.27 $\pm$ 0.03  &  127  \\
26.17  &  -6.14 $\pm$ 0.13  &  20  &  -6.44 $\pm$ 0.09  &  10  &  -6.44 $\pm$ 0.09  &  10  \\
26.82  &  -6.95 $\pm$ 0.35  &  3  &  -7.43 $\pm$ 0.28  &  1  &  -7.13 $\pm$ 0.20  &  2  \\
\hline
\multicolumn{7}{c}{z = 3.6 $-$ 4.0 } \\
25.57  &  -5.09 $\pm$ 0.05  &  172  &  -5.39 $\pm$ 0.03  &  87  &  -5.39 $\pm$ 0.03  &  85  \\
26.22  &  -6.42 $\pm$ 0.19  &  10  &  -6.57 $\pm$ 0.11  &  7  &  -6.94 $\pm$ 0.16  &  3  \\
26.86  &  -7.41 $\pm$ 0.28  &  1  &  $-$ & $-$  &  -7.41 $\pm$ 0.28  &  1  \\
\hline
\hline
\end{tabular}
\end{table*}
\renewcommand{\thetable}{\arabic{table}}

\begin{table*}
\centering
\caption{Estimates of the SFR functions, $\Phi=dN/(d\log SFR\, dV) [\hbox{Mpc}^{-3}]$, at redshifts of up to
$\simeq 3$ for SFGs, RQ AGN and the sum of the two populations. Errors are the quadratic sum of Poisson uncertainties and sample variance. On the right of the SFR functions we give the
effective number, N, of sources in each bin.}
\begin{tabular}{ccccccc}
\hline
\hline
$\log SFR$ & $\log(\Phi_{\rm SFG + RQ AGN})$ & N$_{\rm SFG + RQ AGN}$ & $\log(\Phi_{\rm SFG})$ & N$_{\rm SFG}$ & $\log(\Phi_{\rm RQ\, AGN})$ & N$_{\rm RQ\,AGN}$  \\
$[\hbox{M}_{\odot}\,\hbox{yr}^{-1}]$ & $[\hbox{Mpc}^{-3}\,\hbox{dex}^{-1}]$ & & $[\hbox{Mpc}^{-3}\,\hbox{dex}^{-1}]$ & &
$[\hbox{Mpc}^{-3}\,\hbox{dex}^{-1}]$ & \\
\hline
\hline
\multicolumn{7}{c}{z = 0.05$-$ 0.2 } \\
-1.80  &  -2.45 $\pm$ 0.06  &  1384.3  &  -2.51 $\pm$ 0.02  &  1209.5  &  -3.35 $\pm$ 0.06  &  174.8  \\
-1.10  &  -2.47 $\pm$ 0.21  &  1320.9  &  -2.52 $\pm$ 0.02  &  1182.0  &  -3.45 $\pm$ 0.21  &  138.9  \\
-0.40  &  -2.55 $\pm$ 0.26  &  1095.3  &  -2.57 $\pm$ 0.05  &  1053.4  &  -3.97 $\pm$ 0.25  &  41.9  \\
0.30  &  -2.83 $\pm$ 0.31  &  569.5  &  -2.85 $\pm$ 0.27  &  552.8  &  -4.37 $\pm$ 0.16  &  16.7  \\
1.00  &  -3.82 $\pm$ 0.78  &  59.1  &  -3.88 $\pm$ 0.69  &  51.6  &  -4.72 $\pm$ 0.36  &  7.5  \\
1.70  &  -5.03 $\pm$ 0.79  &  3.6  &  -5.38 $\pm$ 0.48  &  1.6  &  -5.29 $\pm$ 0.63  &  2.0  \\
2.40  &  -5.54 $\pm$ 0.69  &  1.1  &  -5.89 $\pm$ 0.60  &  0.5  &  -5.79 $\pm$ 0.33  &  0.6  \\
\hline
\multicolumn{7}{c}{z = 0.2 $-$ 0.4 } \\
0.50  &  -2.97 $\pm$ 0.15  &  2704.0  &  -2.99 $\pm$ 0.13  &  2560.0  &  -4.24 $\pm$ 0.06  &  144.0  \\
1.25  &  -3.73 $\pm$ 0.95  &  460.3  &  -3.76 $\pm$ 0.72  &  434.7  &  -4.99 $\pm$ 0.62  &  25.6  \\
2.00  &  -5.42 $\pm$ 1.26  &  9.7  &  -5.49 $\pm$ 1.10  &  8.1  &  -6.20 $\pm$ 0.62  &  1.6  \\
2.75  &  -5.97 $\pm$ 0.91  &  2.6  &  -6.05 $\pm$ 0.64  &  2.2  &  -6.75 $\pm$ 0.64  &  0.4  \\
\hline
\multicolumn{7}{c}{z = 0.4 $-$ 0.6 } \\
1.75  &  -4.36 $\pm$ 1.19  &  271.5  &  -4.43 $\pm$ 0.99  &  231.3  &  -5.19 $\pm$ 0.65  &  40.2  \\
2.60  &  -6.28 $\pm$ 0.89  &  3.3  &  -6.56 $\pm$ 0.65  &  1.7  &  -6.60 $\pm$ 0.61  &  1.6  \\
3.45  &  -6.88 $\pm$ 0.93  &  0.8  &  -7.18 $\pm$ 0.71  &  0.4  &  -7.18 $\pm$ 0.60  &  0.4  \\
\hline
\multicolumn{7}{c}{z = 0.6 $-$ 0.8 } \\
1.25  &  -3.39 $\pm$ 0.29  &  3474.9  &  -3.41 $\pm$ 0.28  &  3332.7  &  -4.78 $\pm$ 0.07  &  142.2  \\
2.00  &  -4.46 $\pm$ 1.53  &  295.2  &  -4.52 $\pm$ 1.37  &  258.7  &  -5.37 $\pm$ 0.67  &  36.5  \\
2.75  &  -6.01 $\pm$ 0.63  &  8.3  &  -6.16 $\pm$ 0.45  &  5.9  &  -6.55 $\pm$ 0.44  &  2.4  \\
3.50  &  -6.58 $\pm$ 0.92  &  2.2  &  -6.72 $\pm$ 0.65  &  1.6  &  -7.12 $\pm$ 0.65  &  0.6  \\
4.25  &  -6.90 $\pm$ 0.03  &  1.1  &  -7.05 $\pm$ 0.02  &  0.8  &  -7.44 $\pm$ 0.02  &  0.3  \\
\hline
\multicolumn{7}{c}{z = 0.8 $-$ 1.0 } \\
2.30  &  -4.61 $\pm$ 0.71  &  298.1  &  -4.66 $\pm$ 0.49  &  263.4  &  -5.54 $\pm$ 0.51  &  34.7  \\
3.10  &  -6.33 $\pm$ 0.56  &  5.7  &  -6.42 $\pm$ 0.50  &  4.6  &  -7.02 $\pm$ 0.26  &  1.1  \\
3.90  &  -6.85 $\pm$ 0.89  &  1.7  &  -6.97 $\pm$ 0.66  &  1.3  &  -7.46 $\pm$ 0.59  &  0.4  \\
\hline
\multicolumn{7}{c}{z = 1.0 $-$ 1.2 } \\
2.30  &  -4.42 $\pm$ 1.07  &  554.2  &  -4.45 $\pm$ 0.89  &  511.6  &  -5.53 $\pm$ 0.59  &  42.6  \\
3.10  &  -6.27 $\pm$ 0.74  &  7.9  &  -6.43 $\pm$ 0.57  &  5.4  &  -6.76 $\pm$ 0.47  &  2.5  \\
3.90  &  -7.03 $\pm$ 0.91  &  1.4  &  -7.34 $\pm$ 0.58  &  0.7  &  -7.32 $\pm$ 0.71  &  0.7  \\
\hline
\multicolumn{7}{c}{z = 1.2 $-$ 1.4 } \\
1.75  &  -4.09 $\pm$ 0.29  &  1411.4  &  -4.13 $\pm$ 0.22  &  1280.4  &  -5.12 $\pm$ 0.19  &  131.0  \\
2.60  &  -5.00 $\pm$ 1.00  &  175.1  &  -5.05 $\pm$ 0.77  &  153.9  &  -5.91 $\pm$ 0.64  &  21.2  \\
3.45  &  -6.14 $\pm$ 0.80  &  12.5  &  -6.19 $\pm$ 0.63  &  11.2  &  -7.11 $\pm$ 0.50  &  1.3  \\
4.30  &  -6.19 $\pm$ 0.68  &  11.3  &  -6.20 $\pm$ 0.02  &  10.9  &  -7.65 $\pm$ 0.68  &  0.4  \\
\hline
\multicolumn{7}{c}{z = 1.4 $-$ 1.6 } \\
2.30  &  -4.53 $\pm$ 0.73  &  512.9  &  -4.61 $\pm$ 0.59  &  432.4  &  -5.34 $\pm$ 0.43  &  80.5  \\
3.10  &  -6.00 $\pm$ 1.14  &  17.6  &  -6.17 $\pm$ 0.90  &  11.9  &  -6.49 $\pm$ 0.69  &  5.7  \\
3.90  &  -7.12 $\pm$ 0.86  &  1.4  &  -7.42 $\pm$ 0.61  &  0.7  &  -7.43 $\pm$ 0.61  &  0.7  \\
\hline
\multicolumn{7}{c}{z = 1.6 $-$ 1.8 } \\
2.00  &  -3.94 $\pm$ 0.25  &  2366.8  &  -3.99 $\pm$ 0.19  &  2137.7  &  -4.96 $\pm$ 0.17  &  229.1  \\
2.90  &  -4.91 $\pm$ 0.93  &  257.7  &  -4.97 $\pm$ 0.61  &  223.8  &  -5.79 $\pm$ 0.69  &  33.9  \\
3.80  &  -6.27 $\pm$ 0.84  &  11.3  &  -6.29 $\pm$ 0.59  &  10.7  &  -7.55 $\pm$ 0.59  &  0.6  \\
\hline
\multicolumn{7}{c}{z = 1.8 $-$ 2.0 } \\
2.00  &  -3.93 $\pm$ 0.21  &  2536.0  &  -3.97 $\pm$ 0.19  &  2314.9  &  -4.99 $\pm$ 0.10  &  221.1  \\
2.90  &  -4.82 $\pm$ 1.23  &  328.1  &  -4.91 $\pm$ 1.02  &  265.8  &  -5.54 $\pm$ 0.70  &  62.3  \\
3.80  &  -6.39 $\pm$ 0.83  &  8.7  &  -6.93 $\pm$ 0.55  &  2.5  &  -6.54 $\pm$ 0.62  &  6.2  \\
\hline
\multicolumn{7}{c}{z = 2.0 $-$ 2.4 } \\
2.25  &  -4.22 $\pm$ 0.25  &  2819.4  &  -4.26 $\pm$ 0.20  &  2568.4  &  -5.27 $\pm$ 0.15  &  251.0  \\
3.20  &  -5.38 $\pm$ 1.47  &  195.2  &  -5.46 $\pm$ 1.18  &  162.1  &  -6.15 $\pm$ 0.88  &  33.1  \\
4.15  &  -6.69 $\pm$ 0.90  &  9.5  &  -6.85 $\pm$ 0.64  &  6.6  &  -7.21 $\pm$ 0.64  &  2.9  \\
\hline
\multicolumn{7}{c}{z = 2.4 $-$ 2.8 } \\
2.25  &  -4.39 $\pm$ 0.37  &  1914.2  &  -4.43 $\pm$ 0.27  &  1743.8  &  -5.44 $\pm$ 0.25  &  170.4  \\
3.20  &  -5.18 $\pm$ 1.19  &  309.8  &  -5.27 $\pm$ 1.02  &  252.1  &  -5.91 $\pm$ 0.60  &  57.7  \\
4.15  &  -7.16 $\pm$ 0.73  &  3.2  &  -7.34 $\pm$ 0.52  &  2.1  &  -7.64 $\pm$ 0.52  &  1.1  \\
\hline
\multicolumn{7}{c}{z = 2.8 $-$ 3.2 } \\
2.25  &  -4.67 $\pm$ 0.51  &  981.1  &  -4.75 $\pm$ 0.36  &  821.0  &  -5.46 $\pm$ 0.37  &  160.1  \\
3.20  &  -5.03 $\pm$ 0.59  &  435.4  &  -5.16 $\pm$ 0.47  &  319.4  &  -5.60 $\pm$ 0.35  &  116.0  \\
4.15  &  -6.64 $\pm$ 0.74  &  10.6  &  -6.77 $\pm$ 0.52  &  7.8  &  -7.21 $\pm$ 0.52  &  2.8  \\
\hline
\hline
\end{tabular}
\label{tab:SFRf}
\end{table*}

\end{document}